\documentclass[%
 reprint,
superscriptaddress,
 amsmath,amssymb,
 aps,
]{revtex4-2}

\usepackage{graphicx}
\usepackage{dcolumn}
\usepackage{bm}
\usepackage{xr}
\usepackage{array}
\usepackage{enumitem}
\usepackage{amsthm}
\usepackage{xcolor}
\usepackage{amssymb}
\usepackage{hyperref}
\usepackage{mathtools}
\usepackage{booktabs}
\usepackage{makecell}
\makeatletter
\newcommand*{\addFileDependency}[1]{
  \typeout{(#1)}
  \@addtofilelist{#1}
  \IfFileExists{#1}{}{\typeout{No file #1.}}}
\makeatother

\newcommand*{\myexternaldocument}[1]{
    \externaldocument{#1}
    \addFileDependency{#1.tex}
    \addFileDependency{#1.aux}}
\myexternaldocument{MainSM}
\usepackage{hyperref}%

\usepackage{xcolor} 

\newcommand{\rev}[1]{\textcolor{black}{#1}}

\usepackage{amsthm}

\theoremstyle{definition}

\begin{document}

\preprint{APS/123-QED}

\title{Interpretable disorder-promoted synchronization \\ and coherence in coupled laser networks}%

\author{Ana Elisa D. Barioni} 
\author{Arthur N. Montanari}
\affiliation{Center for Network Dynamics, Northwestern University, Evanston, IL 60208}
\affiliation{Department of Physics and Astronomy, Northwestern University, Evanston, IL 60208}

\author{Adilson E. Motter}
\affiliation{Center for Network Dynamics, Northwestern University, Evanston, IL 60208}
\affiliation{Department of Physics and Astronomy, Northwestern University, Evanston, IL 60208}
\affiliation{Department of Engineering Sciences and Applied Mathematics, Northwestern University, Evanston, IL 60208}
\affiliation{Northwestern Institute on Complex Systems, Northwestern University, Evanston, IL 60208}

\date{\today}%
\begin{abstract}
Coupled lasers offer a promising approach to scaling the power output of photonic devices for applications demanding high frequency precision and beam coherence. However, maintaining coherence among lasers remains a fundamental challenge due to desynchronizing instabilities arising from time delay in the optical coupling.
    Here, we depart from the conventional notion that disorder is detrimental to synchronization and instead propose an interpretable mechanism through which heterogeneity in the laser parameters can be harnessed to promote synchronization. 
    Our approach allows stabilization of pre-specified synchronous states that, while abundant, are often unstable in systems of identical lasers.
    The results show that stable synchronization enabling coherence can be frequently achieved by introducing intermediate levels of random mismatches in any of several laser constructive parameters.
    Our results establish a principled framework for enhancing coherence in large laser networks, offering a robust strategy for power scaling in photonic systems. 

\medskip\noindent
Published in \textit{Physical Review Letters} 135, 197401 (2025). DOI: \href{https://doi.org/10.1103/8qvk-hpwb}{
10.1103/8qvk-hpwb}
\end{abstract}

\maketitle

\textit{Introduction}\textemdash Semiconductor lasers are widely used in applications demanding high frequency precision and low noise \cite{chembo2019optoelectronic}, including interferometric sensing  \cite{li2017laser}, spectrography \cite{park2000diode}, optical communication \cite{hillmer2004low}, and cryptography \cite{banerjee2011synchronization}. 
These applications generally require the lasers to operate in \textit{single-mode} regimes in the terahertz gap with a narrow linewidth of the order of 1 kHz. 
In this context, the sensing and communication capabilities depend on the beam output power. Yet, individual lasers have very low power ($\sim$1~mW) \cite{svelto2010principles}, and scaling their power output while maintaining beam coherence is a challenging task.
Coupled laser systems offer a solution by enabling high-power devices with stable frequency and narrow linewidth. A coherent combined beam can, in principle, be achieved through a coupling-mediated synchronization of the individual laser fields.  However, in practice, the lasers exhibit instabilities due to time delays involved in optical feedback \cite{van1998laser}, which gives rise to relaxation oscillations \cite{haken1963frequency,liu2021influence}, low-frequency fluctuations \cite{torcini2006low,vaschenko1998temporal}, and regular pulse packages \cite{ruschel2017chaotic}, \rev{\cite{heil2001dynamics}}. 
These effects result in chaotic behavior and power dropouts that adversely impact synchronization.

Previous studies have shown that tuning laser parameters 
can improve synchronization across different time delays \cite{junges2013characterization,shena2017turbulent,yanchuk2010multiple, rottschafer2007ecm} and coupling schemes \cite{li2013synchronizing,soriano2013complex,shena2017turbulent,ding2019dispersive,hart2019topological,carroll2004dynamics}. 
These studies have focused primarily on arrays of identical lasers \cite{liu2010coherent,ito2024conflict,rogister2004Power,hart2019delayed}, while a few have modeled the detrimental effects posed by imperfections in manufacturing \cite{vladimirov2003synchronization,nair2019almost}. 
Thus, it has been generally assumed that parameter mismatches inhibit synchronization in networks of coupled oscillators \cite{Rodrigues2016kuramoto,kouomou2004effect,hicke2011mismatch}.
However, advances in nonlinear dynamics have shown that heterogeneity can, in fact, enhance synchronization in a range of network systems \cite{nishikawa2016symmetric,brandt2006synchronization,braiman1995taming,zhang2021random}, including power grids \cite{molnar2021asymmetry}, electronic circuits \cite{braiman1995disorder,mallada2015distributed},  and quantum systems \cite{lorch2017quantum}.
In coupled laser arrays, external cavity misalignment\textemdash which introduces heterogeneity in the coupling delays\textemdash has been shown to facilitate phase locking \cite{nair2021using}. Conversely, it has also been shown that frequency detuning reduces coherence in non-delayed laser arrays \cite{pando2024synchronization}. \rev{This contrast raises fundamental questions about which laser classes can benefit from heterogeneities, and how coupling delays interact with parameter disorder to influence synchronization.}

\begin{figure}[b!]
\centering
\includegraphics[width=0.95\linewidth]{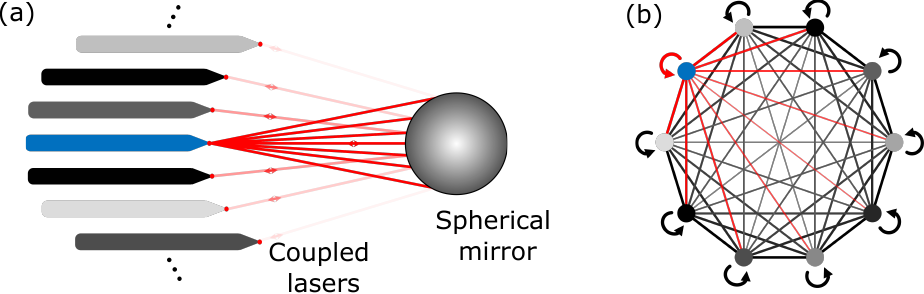}
\caption{Coupled-laser systems. 
(a) Schematic of a laser array coupled through a spherical mirror. The light paths are depicted for the blue laser, showing that the coupling strength decays with inter-laser distance due to mirror diffraction.
(b) Decaying network topology. In both panels, the line thickness represents the coupling strength, and the shades of gray indicate the heterogeneity in laser parameters.}
\label{fig:Schematics}
\end{figure}

In this letter, we establish an {interpretable} mechanism in which disordered heterogeneity enables synchronization in systems of delayed-coupled lasers with arbitrary networks.
The mechanism is interpretable as it enables the {\it stabilization of pre-specified states}, and the heterogeneity is disordered in the sense that this synchronization can be achieved with suitable levels of {\it random} heterogeneity in laser parameters.
We focus on frequency-synchronized states, where lasers share the same frequency and exhibit small phase mismatches, as they yield a coherent beam whose power scales with network size. The method is powerful because, as we show, such states are abundant, but usually unstable in the absence of heterogeneity. 
Our results show that controlled levels of random disorder in the (intrinsic) frequency detuning, as well as other laser parameters, can lead to a transition of the frequency-synchronized states from unstable to stable with high probability.
As a consequence, coupled disordered systems with up to a thousand lasers can achieve a sharper frequency spectrum and higher coherence compared to their homogeneous counterparts.

\textit{Dynamics of delay-coupled laser networks}\textemdash
A network of $M$ single-mode lasers with time-delay coupling  can be modeled by the Lang-Kobayashi (LK) equations \cite{lang1980external,masoller1997implications}:
\begin{equation}
\begin{aligned}
    \dot{r}_j(t)&=\frac{1}{2}\left(G_j-\gamma\right) r_j(t)+ {\kappa_j}\sum_{k=1}^M {A}_{j k} r_k(t-\tau) \cos\Phi_{jk},
    \\
    \dot{\phi}_j(t)&=\frac{\alpha_j}{2}\left(G_j-\gamma\right)+ \omega_j+ {\kappa_j}\sum_{k=1}^M {A}_{j k} \frac{r_k(t-\tau)}{r_j(t)} \sin\Phi_{jk},
    \\
    \dot{N}_j(t)&=J_{0}-\gamma_{n} N_j(t)-G_j r_j^2(t),
\end{aligned}
   \label{LK_eqs_polar}
\end{equation}

\noindent
where $r_j(t)$ and $\phi_j(t)$ are respectively the amplitude and phase of the electric field $E_j(t)=r_j(t)e^{i\phi_j(t)}$ of laser $j$, $N_j(t)$ is the corresponding carrier number, and $\Phi_{jk}=\phi_k(t-\tau)-\phi_j(t)$ denotes the time-delay coupling. Here, $\omega_j$ is the frequency detuning with respect to the lasing mode frequency $\omega_0$ (determined by the internal cavity size), $\alpha_j$ is the linewidth enhancement factor accounting for the phase-amplitude coupling, and $G_j(t)=g [(N_j(t)-N_{0})/(1+sr_j^2(t))]$ is the nonlinear optical gain. The other constructive parameters (and corresponding values assumed throughout) are the gain coefficient $g=1.5\times 10^{-5} \,\text{ns}^{-1}$, gain saturation coefficient $s=10^{-7}$,  cavity loss $\gamma=500 \,\text{ns}^{-1}$, carrier loss rate $\gamma_n=0.5 \,\text{ns}^{-1}$, carrier number at transparency $N_0=1.5\times 10^8$, pump current $J_{0}=g_{p} \gamma_{n}\left(N_{0} + \frac{\gamma}{g}\right)$, and pump gain $g_{p}=2.55$. The tunable parameters are set as $\tau=0.15$~ns, $\omega_j=0$, and $\alpha_j=5$, $\forall j$, unless specified otherwise. These parameter choices are consistent with realistic experimental conditions \cite{kozyreff2001dynamics,liu2008coherent,liu2014nonlinear}. 

\begin{figure*}[t!]
\centering
\includegraphics[width=1\linewidth]{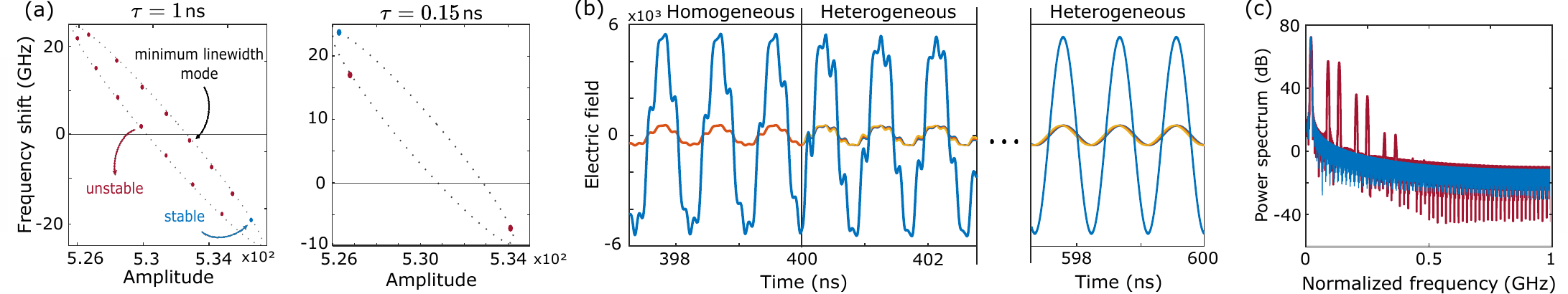}
\caption{Synchronization dynamics and coherence in coupled disordered lasers.
(a)~Spectrum of stationary solutions (amplitude $r^*$ and frequency $\Omega$ pairs) for $\tau = 1$ and $0.15$ ns, showing the multistability of the LK model. Stable and unstable modes are marked in blue and red, respectively.
(b)~Time series of the electric fields as the lasers transition from a homogeneous configuration for $t\in[0,400]$ ns (where  $\omega_j=0$, $\forall j$) to a heterogeneous configuration for $t\in[400,600]$ ns (where $\omega_j$ follows a normal distribution). The small-amplitude signals indicate the imaginary component of the individual fields $E_j$, while the high-amplitude signal represents the combined field $E=\sum_jE_j$. For the homogeneous configuration, the lasers converge to a limit cycle with zero phase shift but distorted waveforms, whereas the heterogeneous configuration outputs sinusoidal waves with constant, but negligible, phase shifts among lasers. \rev{The frequency and amplitude dynamics are included in Fig.~S7~\cite{supplemental_mat}.}
(c) Power spectrum of the combined steady-state field for the homogeneous (red) and heterogeneous (blue) configurations.
The simulations are on the 10-laser decaying network for $\kappa=0.54\, \rm{ns}^{-1}$ and $\tau=0.15$ ns. }
\label{fig:dynamics}
\vspace{-0.3cm}
\end{figure*}

The array of single-mode semiconductor lasers is coupled through a reflector, enabling long-range, global interactions \cite{kozyreff2001dynamics}, as illustrated in Fig.~\ref{fig:Schematics}(a). Pairwise field interactions are modeled as a network [Fig.~\ref{fig:Schematics}(b)], where the adjacency matrix ${A}=(A_{jk})$ characterizes the (possibly weighted and directed) coupling structure, $\kappa_j$ denotes the global coupling strength, and $\tau$ is the time delay related to the optical path length to the grating reflector.
To account for the stronger coupling between spatially closer lasers in an array, we consider the \textit{decaying coupling} function $A_{jk}=(d_x)^n$, where $d_x=0.95$ and $n=|j-k|$ represents the distance between lasers $j$ and $k$ (with periodic boundaries) \cite{nair2021using}. We also investigate the all-to-all coupling (where $A_{jk}=1$, $\forall j,k$) and ring coupling (where $A_{jk}=1$ if $|j-k|=1$ and 0 otherwise). 
We refer to Supplemental Material (SM) \cite{supplemental_mat}, Sec. SI, for a derivation of the LK model from the Maxwell-Bloch equations.

\textit{Synchronization and coherence induced by parameter disorder}\textemdash To generate a coherent beam, all lasers must synchronize to the same frequency with minimal phase spreading to avoid destructive interference. Thus, we focus on stationary synchronous solutions of the form $E_j(t)=r^*_je^{i(\Omega t+\delta^*_j)}$ and $N_j(t)=N^*_j$, where $r_j^*$ is the stationary amplitude and $N_j^*$ is the corresponding carrier number. Here, $\Omega$ is the frequency shift with respect to the lasing mode frequency $\omega_0$\textemdash the ideal operating frequency\textemdash and $\delta^*_j$ are constant phase shifts among lasers. The lasers are said to be coherent if $\delta^*_j\approx 0, \forall j$. Our goal is to establish the conditions under which such synchronous states are stable.

Applying the above ansatz to Eq.~\eqref{LK_eqs_polar} for $\dot r_j=0$, $\dot \phi_j=\Omega $, and $\dot N=0$, leads to the transcendental equations:
\begin{equation}
\begin{aligned}
    0&=\frac{1}{2}\left(G_j-\gamma\right) r_j(t)+ {\kappa_j}\sum_{k=1}^M {A}_{j k} r_k(t-\tau) \cos\Phi_{jk},
    \\
    \Omega&=\frac{\alpha_j}{2}\left(G_j-\gamma\right)+ \omega_j+ {\kappa_j}\sum_{k=1}^M {A}_{j k} \frac{r_k(t-\tau)}{r_j(t)} \sin \Phi_{jk},
    \\
    0&=J_{0}-\gamma_{n} N_j(t)-G_j r_j^2(t).
\end{aligned}
   \label{eq.transceq}
\end{equation}

\noindent
This set of $3M$ equations can be solved for $r_j^*$, $N_j^*$, and $\delta^*_j$,  $\forall j$, along with the corresponding frequency mode $\Omega$ (see SM~\cite{supplemental_mat}, Sec. SII, for details on the computational procedure).
When the lasers have identical parameters, there exist multiple solutions for which \textit{identical synchronization} is achieved (i.e., $E_j(t)=r^*e^{i\Omega t}$ and $ N_j(t)=N^*$, $\forall j$); these solutions fall on an ellipse \cite{flunkert2011delay,soriano2013complex}, as shown in Fig.~\ref{fig:dynamics}(a). Notably, many of these solutions are unstable, and the multistability of the system depends strongly on $\tau$ and $\kappa$ (Fig.~S1 \cite{supplemental_mat}).
Since laser arrays are designed to operate near the lasing mode frequency $\omega_0$ (i.e., when $\Omega \approx 0$) \cite{soriano2013complex,kane2005unlocking}, we focus here on the stabilization of synchronous states associated with the so-called \textit{minimum linewidth mode} $\Omega_{\rm ML}=\min
|\Omega|$. However, as indicated in Fig.~\ref{fig:dynamics}(a), the $\Omega_{\rm ML}$-mode is unstable for \textit{identical} lasers under strong coupling (see also Figs.~S1-S2~\cite{supplemental_mat}). 

When heterogeneity is introduced into laser parameters, the identical synchronization state $E_j(t)=r^*e^{i\Omega t}$ generically ceases to exist and only \textit{frequency-synchronized solutions} $E_j(t)=r^*_j e^{i(\Omega t+\delta^*_j)}$ remain. Notwithstanding, we find that heterogeneous laser arrays can still admit frequency-synchronized states with highly cohesive phases. \rev{The homogeneous-heterogeneous transition is illustrated in Fig.~\ref{fig:dynamics}(b). For a homogeneous configuration, the stationary state $E_j(t)=r^*e^{i\Omega_{\rm ML}t}$ is unstable, causing the lasers to converge to a multimodal limit cycle, as shown for $t\in[397, 400]$. After switching to a suitable heterogeneous configuration,  the synchronous state $E_j(t)=r_j^*e^{i(\Omega_{\rm ML}t+\delta^*_j)}$ becomes stable. Thus, after a desynchronized transient, the system asymptotically converges to this frequency-synchronized state, as shown for $t\in[597, 600]$. Crucially,} the heterogeneous solution is highly coherent given that $\delta^*_j\approx 0$, $\forall j$.
As a result, its combined electric field has a frequency spectrum with a single sharp peak, which contrasts with the multi-peak spectrum in the homogeneous case [Fig.~\ref{fig:dynamics}(c)].

\textit{Interpretable mechanism for disorder-promoted synchronization}\textemdash
To analyze the stability of delayed-coupled systems and characterize their dependence on disorder, we express Eq.~\eqref{LK_eqs_polar} as follows:
\begin{equation}
\dot{\textbf{x}}_j(t) = \textbf{f}_j\big(\textbf{x}_j(t)\big) + {\kappa}_j \sum_{k=1}^M A_{jk} \textbf{h}\big(\textbf{x}_j(t),\textbf{x}_k(t-\tau)\big),
\label{Dyn_LK_Eq}
\end{equation}

\noindent where $\textbf{x}_j=(r_j,\phi_j,N_j)^\top$ is the state of laser $j$, $\textbf{f}_j$ is the vector field describing the uncoupled dynamics, and $\textbf{h}$ is the coupling function. 
To derive the variational equation, we introduce $\boldsymbol{\eta}_j(t)=\big(\delta r_j(t),\delta \phi_j(t),\delta N_j(t)\big)^{\top}$ as the vector of small deviations from the stationary state $\textbf{x}_j^*=(r_j^*,\Omega t+\delta^*_j,N_j^*)^\top$, and define the full perturbation vector $\boldsymbol{\eta}(t)=(\boldsymbol{\eta}_1^\top,\hdots,\boldsymbol{\eta}_M^\top)^\top$. Linearizing around $\textbf{x}^*$ leads to
\begin{equation}
\begin{split}
    \dot{\boldsymbol{\eta}}_j(t) = \operatorname{D}_{\textbf{x}_j}^{(0)}\textbf{f}_j(\textbf{x}^*_j)\boldsymbol{\eta}_j(t) +  \kappa_j d_j\operatorname{D}_{\textbf{x}_j}^{(0)}\textbf{h}_j(\textbf{x}^*)\boldsymbol{\eta}_j(t) \\
+ {\kappa}_j \sum_{k=1}^M A_{jk} \operatorname{D}_{\textbf{x}_k}^{(\tau)}\textbf{h}_j(\textbf{x}^*_j,\textbf{x}^*_k)\boldsymbol{\eta}_k(t-\tau),
\end{split}
\label{Var_LK_Eq}
\end{equation}

\noindent 
where $d_j=\sum_kA_{jk}$ is the node indegree and $\operatorname{D}_{\textbf{x}_j}^{(\tau)}$ denotes the Jacobian matrix of a vector field with respect to the delayed state $\textbf{x}_j(t-\tau)$; accordingly, $\operatorname{D}_{\textbf{x}_j}^{(0)}$ denotes the Jacobian with respect to $\textbf{x}_j(t)$.
Equation~\eqref{Var_LK_Eq} can be organized in matrix form as
\begin{equation}
\begin{aligned}
    \dot{\boldsymbol{\eta}}(t) = {J}_1 \, \boldsymbol{\eta}(t) + {J}_2
 \, \boldsymbol{\eta}(t-\tau),
\end{aligned}
   \label{VariationalEqs}
\end{equation}
\noindent
where  ${J}_1$ and ${J}_2$ are time-independent matrices (reported in SM~\cite{supplemental_mat}, Sec. SII). Equation~\eqref{VariationalEqs} admits a solution of the form $\boldsymbol{\eta}(t)=\boldsymbol{\eta}(0)e^{\lambda_\ell t}$, whose stability exponents $\lambda_\ell$ are determined by the  characteristic equation \cite{bellen2013numerical,richard2003time}
\begin{equation}
\begin{aligned}
    \det\left({J}_1 + {J}_2 \, e^{-\lambda_\ell \tau} - \lambda_\ell {I}_{3M}\right) = 0,
\end{aligned}
   \label{CharacteristicEq}
\end{equation}

\noindent
where $I_{3M}$ denotes the identity matrix of order $3M$. 
Note that, for time-delay systems, there can be infinite exponents $\lambda_\ell$ satisfying Eq.~\eqref{CharacteristicEq}. The synchronization stability is ultimately determined by the largest Lyapunov exponent $\lambda_{\rm max}=\operatorname{max}_\ell\{\operatorname{Re}(\lambda_\ell)\}$, which can be numerically estimated as described in SM~\cite{supplemental_mat}, Sec. SII.

\rev{In the idealized case of identical oscillators, synchronization stability can be analyzed using the master stability function (MSF) formalism for delayed-coupled systems \cite{choe2010controlling, flunkert2010synchronizing}. Assuming that all nodes share the same in-degree (i.e., $d_j=d$, $\forall j$), Eq.~\eqref{Var_LK_Eq} decouples into $M$ independent eigenmodes:}
\begin{equation}
\rev{\dot{\boldsymbol{\xi}}_j(t) =
\underbrace{\left[\operatorname{D}^{(0)}\!\textbf{f}+\kappa d\operatorname{D}^{(0)}\textbf{h}\right]\boldsymbol{\xi}_j(t) + \kappa \nu \operatorname{D}^{(\tau)}\! \textbf{h} \,\boldsymbol{\xi}_j(t-\tau)}_{\Theta(\mathbf x)},}
\end{equation}
\noindent
\rev{where $\nu$ represents the eigenvalues of $A$. However, in the presence of heterogeneity, this reduction fails due to the persistence of inter-mode coupling terms $\boldsymbol{\Delta}^{(\tau)} (\textbf{x})$ in the variational equations \cite{sugitani2021synchronizing}:
$
\dot{\boldsymbol{\xi}}_j(t)=\Theta(\mathbf x)+\sum_k^M\boldsymbol{\Delta}_{jk}^{(0)}\boldsymbol{\xi}_k(t)+\sum_k^M\boldsymbol{\Delta}_{jk}^{(\tau)}\boldsymbol{\xi}_k(t-\tau).
$
This inter-mode coupling acts as the mechanism underlying the stabilizing effect of heterogeneity. As we show next, stability can emerge through random mismatches, without requiring fine-tuning or optimization.}

\begin{figure}[t]
\centering
\includegraphics[width=1\linewidth]{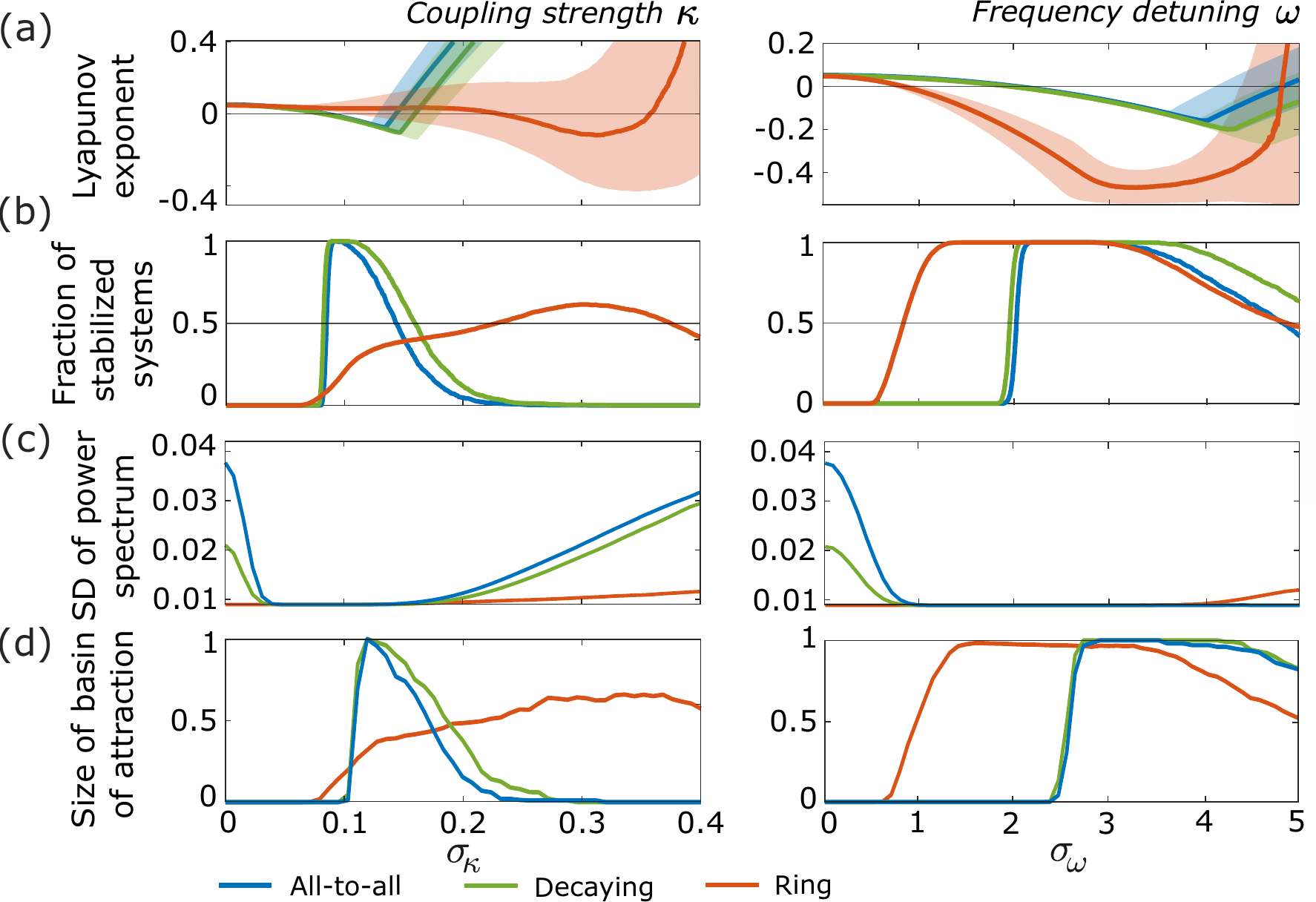}
\caption{Synchronization stability for varying disorder levels in laser arrays.
(a) Lyapunov exponent $\lambda_{\rm max}$ versus disorder level $\sigma_p$ in coupling strength (left) and frequency detuning (right) for three network topologies with 10 lasers: all-to-all (blue), decaying (green), and ring (orange). 
The lines indicate the median across 1,000 realizations of parameter disorder, while shaded areas indicate the first and third quartiles.
(b)  Fraction of realizations that stabilize the frequency-synchronized state. 
(c)  Standard deviation of the power spectrum of the combined steady-state field $E(t)$, averaged over random initial conditions. The standard deviation measures the frequency bandwidth of $E(t)$, as illustrated in Fig.~\ref{fig:dynamics}(c).
(d) Average size of the basin of attraction.
See SM~\cite{supplemental_mat}, Sec. SIII, for numerical analysis.}
\label{fig:results_random}
\vspace{-0.4cm}
\end{figure}

Figure~\ref{fig:results_random} characterizes the relationship between synchronizability and disorder for different laser parameters: $\kappa_j$ and $\omega_j$ (results for $\alpha_j$ are also reported separately in Fig.~S5~\cite{supplemental_mat}). Disorder is systematically introduced by perturbing laser parameters according to $p_j=p_{\rm hom}+\delta p_j$, where $p\in\{\kappa,\omega,\alpha\}$ denotes the disordered parameter, $p_{\rm hom}$ is the homogeneous baseline, and $\delta p_j\sim\mathcal N(0,\sigma_p^2)$ is a Gaussian perturbation with zero mean and standard deviation $\sigma_p$. To preserve the mean, we enforce the constraint $\sum_j \delta p_j=0$, which holds as $M\rightarrow \infty$.
For each disorder realization $\delta p$, we start the simulations with a homogeneous set of parameters (at $\sigma_p=0$), where the identical state $E_j(t)=r^*e^{i\Omega_{\rm ML} t}$ is unstable. By increasing $\sigma_p$, this state ceases to exist and we track instead the stability of the closest synchronous state $E_j(t)=r_j^*e^{i(\Omega_{\rm ML}t + \delta^*_j)}$ via numerical continuation (SM~\cite{supplemental_mat}, Sec. SII).

Figure~\ref{fig:results_random}(a) shows that the median $\lambda_{\rm max}$ decreases as a function of $\sigma_p$, leading to a transition point where the synchronous state is stabilized. 
Importantly, over a wide range of $\sigma_p$, there is a high probability that almost all realizations of parameter disorder can induce stability [Fig.~\ref{fig:results_random}(b)] Such a result is consistent for different parameters and network topologies, including irregular ones (Fig.~S8~\cite{supplemental_mat}).  For instance, ring networks with disordered $\omega_j$ exhibit a robust range $\sigma_{\omega}\in[1.5, 3.2]$ where 100\% of the systems stabilize. \rev{Within the networks studied, the stabilization effect is especially pronounced in sparser topologies, which is consistent with the desynchronizing effect of delayed feedback for large coupling.}

\rev{From a design perspective, parameter disorder substantially widens stability margins. For instance, for the ring networks, disorder in $\omega_j$ extends the range of coupling strengths $\kappa$ and time delays $\tau$ for stable synchronization, respectively, by 23\% and 89\% (Fig.~S9~\cite{supplemental_mat}).
 While one could question whether disorder would inadvertently cause large phase mismatches, our findings confirm that coherence remains persistently high [as anticipated in Fig.~\ref{fig:dynamics}\rev{(b)-(c)}].} Figure~\ref{fig:results_random}(c) shows that the combined field $E=\sum_j E_j$ exhibits a sharp frequency spectrum with a standard deviation near 0.01 over the same range of $\sigma_p$ corresponding to a stable synchronous state.  
 As $\sigma_p$ increases, phase shifts gradually grow, thereby reducing the combined beam coherence. 
 For comparison, we consider in SM~\cite{supplemental_mat}, Sec. SIII, the alternative\textemdash but outperformed\textemdash strategy of applying a constrained type of disorder that preserves the existence of the identical synchronization state.

The beneficial role of disorder is further highlighted in Fig.~\ref{fig:results_random}(d). A suitable choice of disorder not only promotes linear stability $\lambda_{\rm max}$, but also expands the basin of attraction \textit{over the same ranges of} $\sigma_p$. 
Notably, despite the multistable nature of the system, the basin of attraction of the $\Omega_{\rm ML}$-mode dominates the state space. \rev{This dominance occurs because, as shown in Figs.~S3-S4~\cite{supplemental_mat}, all other modes either remain unstable or vanish through bifurcations upon increasing disorder.}
\rev{Given that stronger phase-amplitude coupling (controlled by $\alpha$) is generally associated with increased multistability \cite{bohm2015amplitude}, it is interesting that the basin of attraction of the $\Omega_{\rm ML}$-mode can be expanded by introducing heterogeneity in $\alpha$ rather than by tuning its average  (Fig.~S5~\cite{supplemental_mat}).}

\begin{figure}[t]
\centering
\includegraphics[width=1\linewidth]{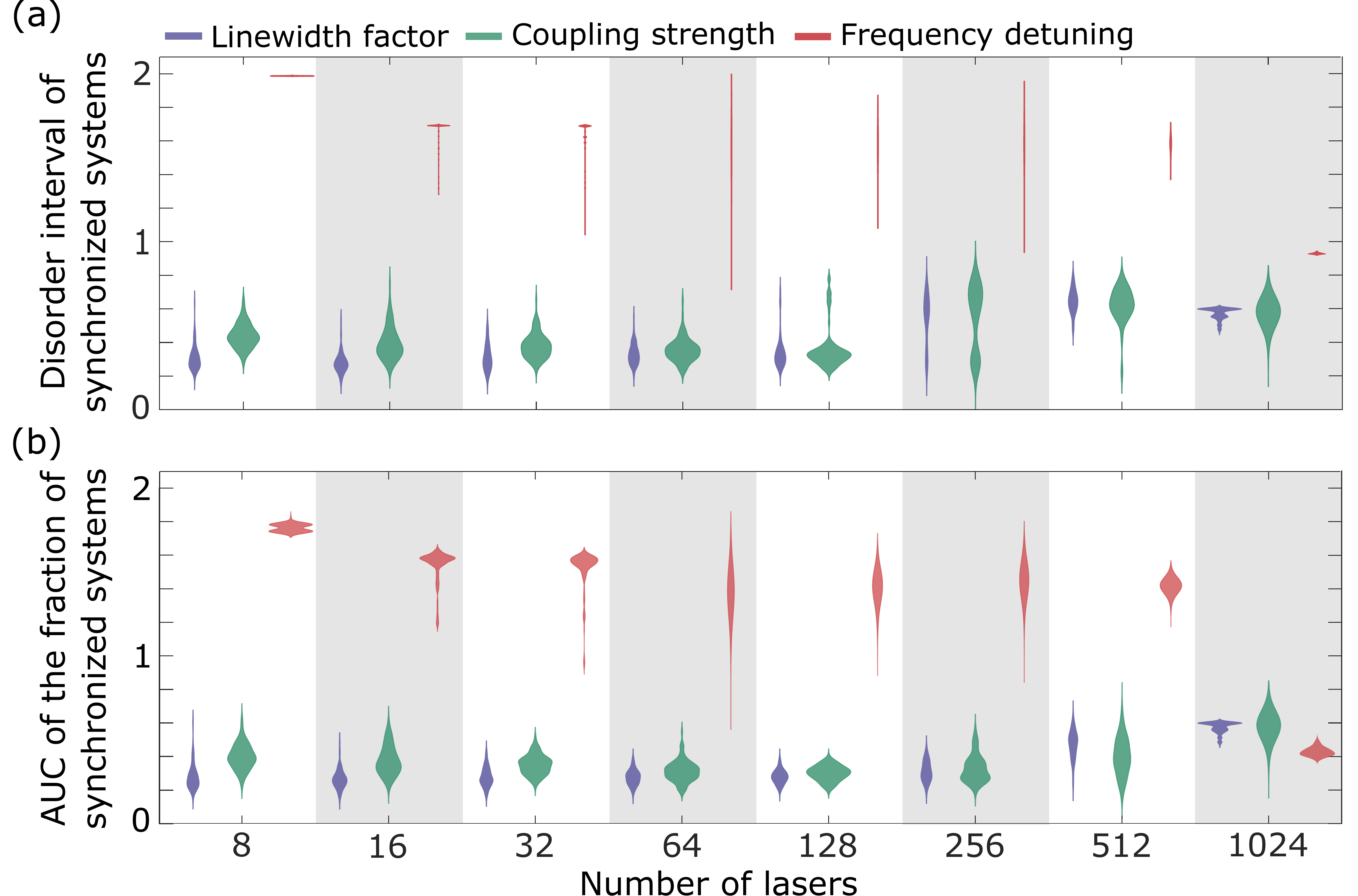}
\caption{Scalability analysis for large laser arrays. Violin plots showing the impact of increasing array sizes on synchronizability. (a) Width of the interval of $\sigma_p$ for which at least 50\% of all disorder realizations lead to a frequency-synchronized system. (b) Area under the curve (AUC) representing the fraction of synchronized systems as a function of $\sigma_p$. These two measures are based on the curves displayed in Fig.~\ref{fig:results_random}(b). In both panels, the distributions include 100 realizations of disorder applied to $\alpha$, $\omega$, and $\kappa$ in a decaying network. See SM~\cite{supplemental_mat}, Sec. SIII, for simulation details.}
\label{fig:net_scaled}
\vspace{-0.5cm}
\end{figure}

Figure~\ref{fig:net_scaled} further demonstrates that our mechanisms for designing disordered laser systems scale well to arrays of up to a thousand lasers. 
Both the fraction of synchronized systems and the range of disorder levels promoting synchronization remain largely consistent across all array sizes (except for a sharp decrease at $M=1024$ for the $\omega$ parameter).
The SM~\cite{supplemental_mat}, Sec.~SIV, characterizes the stability landscape for 3-laser networks, providing an insight into why almost all random realizations of disorder have a high probability of stabilizing synchronization.

\textit{Critical role of time-delay}\textemdash The presence of time delay is a key factor in our proposed framework for disorder-promoted synchronization. 
In the special case of an LK model without delays ($\tau=0$), parameter disorder leads instead to instability (see SM~\cite{supplemental_mat}, Sec. SV). This observation raises the question of whether time delay generally implies disorder-promoted synchronization in other classes of laser systems. 
To address this, we first extend our analysis to \rev{another non-delayed laser model: the \textit{laser rate equations} (LRE), which provide a general macroscopic description for various laser physics (e.g., solid-state and gas lasers) \cite{rogister2004Power,pando2024synchronization}.} 
We observe that disorder impairs synchronizability in the (non-delayed) LRE model, consistent with our findings for the non-delayed LK model and the conclusions drawn in Ref.~\cite{pando2024synchronization}. 
In light of these results, we have tested the impact of incorporating time delay into the LRE model and observed that disorder still inhibits synchronization. Although this may seem counterintuitive\rev{\textemdash given that both the LK equations and LRE are class-B laser models \cite{paoli1988statistical}\textemdash our MSF analysis} reveals that time delay cannot drive a stability transition in the LRE model (SM~\cite{supplemental_mat}, Sec.~SV).
Ultimately, this difference arises from the distinct nature of the two systems: lasers \rev{described by the LRE} lack phase-amplitude coupling, whereas the LK model incorporates the linewidth factor $\alpha$ observed in semiconductor lasers.

Overall, these results show that disorder-promoted synchronization emerges from an intricate interplay between the intrinsic dynamics of the lasers, coupling structure, and time delays. 
\rev{To further understand the influence of the dynamical degrees of freedom, we also examined reduced (class A) and extended (class C) versions of the LRE and LK models, highlighting the decisive role of the model under consideration  (SM~\cite{supplemental_mat}, Sec.~SVI).}

\textit{Conclusion}\textemdash Coupled-laser systems are a promising design for high-power photonics, but coherence is often disrupted due to instabilities caused by time delays. Here, we demonstrated that heterogeneities in semiconductor lasers can stabilize desired synchronous states. \rev{This stabilization mechanism scales to arrays with more than a thousand lasers\textemdash exceeding the current experimental state of the art \cite{hillmer2004low,kao2016phase,pando2024synchronization}\textemdash and may be further extended through modular design or mean-field coupling \cite{balzer2024canard}.}
We examined explicitly the key parameters $\alpha$, $\omega$, and $\kappa$ due to their relevance in laser design: $\alpha$ relates to refractive index, $\omega$ reflects cavity size variations, and $\kappa$ depends on external components. The significance of heterogeneity in $\omega$, for example, is the subject of recent and ongoing research~\cite{ocampo2025frequency,ye2025disorder}. \rev{Existing experimental techniques can be directly used to validate our results by introducing heterogeneities into $\omega$ or $\kappa$ through spatial light modulators inserted within the internal cavity \cite{ngcobo2013digital,pando2024synchronization}. Beyond frequency synchronization, our proposed mechanisms may also support other collective behaviors in laser systems, including chimera states \cite{bohm2015amplitude}, 
canard cascades \cite{balzer2024canard}, and crowd synchronization \cite{zamora2010crowd,mahler2020experimental}.
In multi-mode systems, phase-shift disorder can increase the number of simultaneous lasing modes even in a single laser \cite{eliezer2022controlling}.
We expect our theoretical results to pave the way for a new generation of laser systems with configurable parameters and provide a foundation for harnessing synchronization in other complex physical and biophysical systems where delays are unavoidable.}


\textit{Acknowledgements}\textemdash The authors thank R. Roy for insightful discussions. 
This research was supported by the U.S.\ Office of Naval Research under grant No.\ N00014-22-1-2200.

\rev{\textit{Data availability}\textemdash The code and data that support the findings of this article are openly available \cite{Data_avail}.}


\let\oldaddcontentsline\addcontentsline
\renewcommand{\addcontentsline}[3]{}

\let\addcontentsline\oldaddcontentsline
\clearpage
\onecolumngrid
\renewcommand{\thepage}{\arabic{page}}
\setcounter{page}{1}
\renewcommand{\thefigure}{S\arabic{figure}}
\setcounter{figure}{0}
\setcounter{secnumdepth}{3}
\renewcommand{\thesection}{S\Roman{section}}
\setcounter{section}{0}
\renewcommand{\theequation}{S\arabic{equation}}
\setcounter{equation}{0}
\begin{center}
    \textbf{\large SUPPLEMENTAL MATERIAL \\[0.2cm]
    Interpretable disorder-promoted synchronization and coherence in coupled laser networks}
\end{center}

\tableofcontents




\date{\today}

\section{Derivation, assumptions,  and multistability analysis of the Lang-Kobayashi model}
\label{sec.LKmodel}
For completeness, we derive the LK model to facilitate the understanding of its stability and dynamical properties. We start from the Maxwell-Bloch equations, which provide a fundamental framework to describe the interaction between light and matter, particularly in systems where the gain medium is modeled as a collection of two-level atoms. 
The state variables are \cite{agrawal2013semiconductor,ohtsubo2017semiconductor}:
i)~the electric field $E(t)=E_0(t) e^{i(\omega+\Omega) t} + 
\bar E_0(t) e^{-i(\omega+\Omega) t}$, where $E_0(t)$ is a slowly-varying amplitude;
ii) the polarization $P(t)=P_0(t) e^{-i \omega t} + \bar P_0(t) e^{i \omega t}$, representing the induced dipole moment of the gain medium; and iii) the carrier number $N(t)$ measuring the difference between the populations in the upper and lower energy levels. In our notation, the overbar indicates complex conjugation.
The dynamical equations governing those variables are
\begin{subequations} \label{eq:Max-Bloch}
\begin{align}
\label{eq:Max-Bloch_E}
& \frac{\partial^2 E(t)}{\partial z^2}-\frac{1}{c^2} \frac{\partial^2 E(t)}{\partial t^2}=\frac{1}{\epsilon_0 c^2} \frac{\partial^2 P(t)}{\partial t^2}, \\
\label{eq:Max-Bloch_P}
& \frac{d P_0}{d t}+\gamma_{\perp} P_0=i \hbar E_0 (N-N_0), \\
\label{eq:Max-Bloch_N}
& \frac{d N}{d t}=-\gamma_nN-2 g\left(\bar{E_0} P_0+E_0 \bar{P_0}\right),
\end{align}    
\end{subequations}

\noindent
where $\gamma_n$ is the carrier loss rate, $\gamma_\perp$ is the polarization dephasing rate, $c$ is the speed of light, $\epsilon_0$ is the permittivity of free space, and $\hbar$ is the  reduced Plank constant.

To derive the laser equations (1), we first simplify the electric field equation \eqref{eq:Max-Bloch_E} by assuming a slowly time-varying envelope such that $\frac{\partial^2 E}{\partial z^2} \approx 0$ and $\frac{\partial^2 E}{\partial t^2} \approx 2 i (\omega + \Omega) \frac{\partial E_0}{\partial t}$. Thus, the wave equation \eqref{eq:Max-Bloch_E} reduces to the ordinary differential equation $\dot E_0+i(\omega+\Omega) E_0+ \frac{\gamma}{2} E_0=i \frac{1}{2 \epsilon_0} P_0,$ where the additional term on the LHS accounts for losses in the cavity medium, with $\gamma$ representing the loss rate. Next, we assume that the polarization response $P_0 (t)$ follows the electric field envelope $E_0(t)$ adiabatically: $P_0 \approx \frac{i \hbar}{\gamma_{\perp}} (N-N_0)E_0$. We then substitute $P_0$ into Eqs.~\eqref{eq:Max-Bloch_E} and \eqref{eq:Max-Bloch_N}, and define the effective gain as $G(N)=-\frac{ g}{\epsilon_0 \gamma_\perp}(N-N_0)$. Finally, normalizing the fundamental constants to unity, and including both the pump current $J_0$ driving the carrier number and the gain saturation effect (i.e., $G(N,E)=g \frac{N-N_{0}}{1+s\left|E(t)\right|^2}$), the laser equations can be expressed as 
\begin{subequations} \label{eq:LKderiv2}
\begin{align}
\label{eq:LKderiv_E2}
    \dot E_0 e^{i \Omega t} &= \frac{1}{2}\left(G(N,E)-\gamma\right) E_0e^{i \Omega t}-i\omega E_0 e^{i \Omega t},\\
    \label{eq:LKderiv_N2}
    \dot N &=J_0-\gamma_nN-G(N,E) \left|E_0\right|^2.
\end{align}    
\end{subequations}

\begin{figure}[t]
    \centering
    \includegraphics[width=0.9\linewidth]{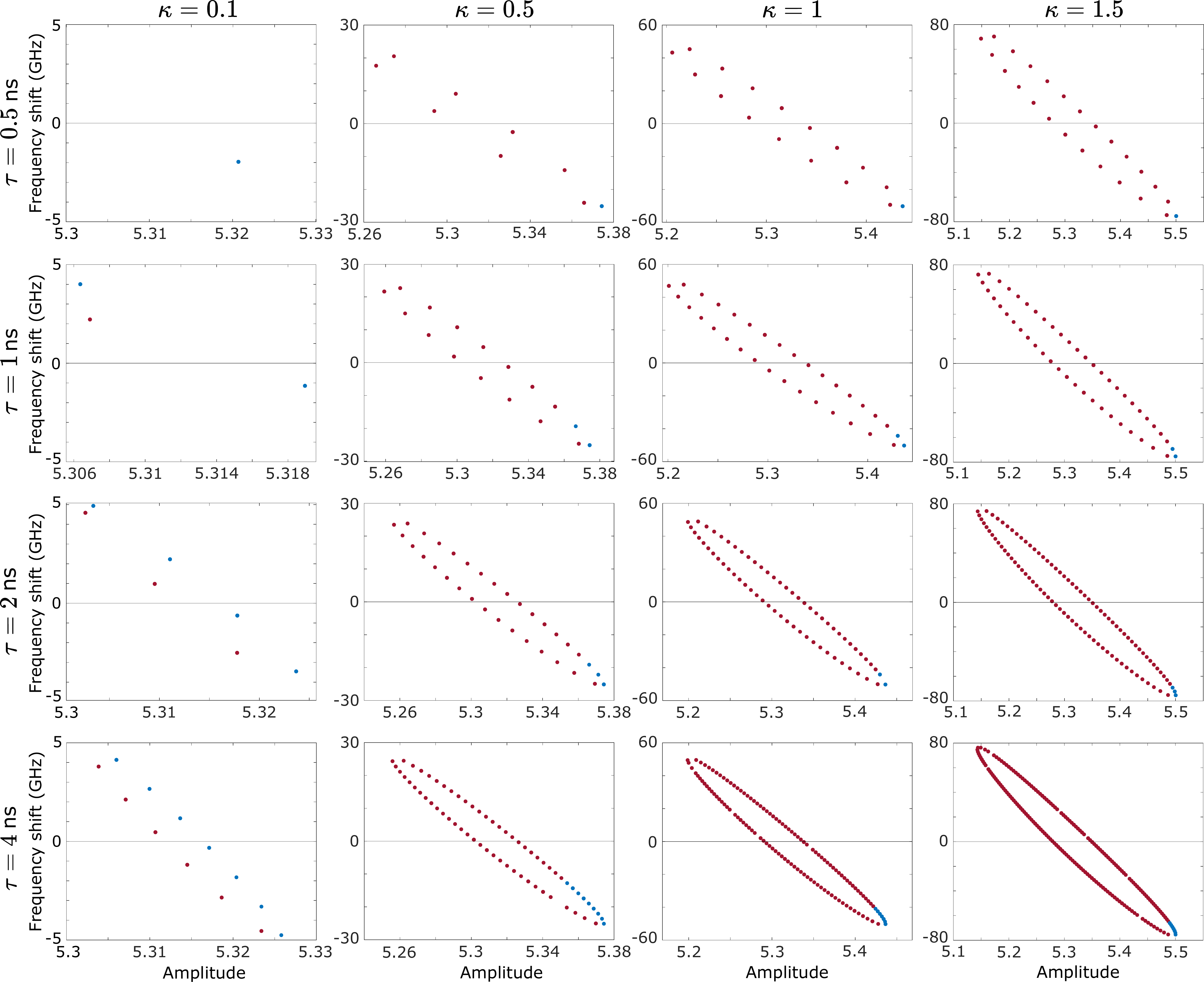}
    \caption{Multistability of the LK model for a homogeneous laser array. Stationary solutions (amplitude $r^*$ and frequency shifts $\Omega$) for increasing time delay $\tau$ (top to bottom panels) and coupling strength $\kappa$ (left to right panels) for a decaying network of 10 lasers. Stable and unstable modes are marked in blue and red, respectively. Note that the number of stationary solutions increases as a function of $\tau$ and $\kappa$.}    \label{fig:multistability_delay}
\end{figure}

The original LK model \cite{lang1980external} for a \textit{single} laser introduces a delayed optical feedback term to the previous equations in order to account for the optical interference caused by the beam reflection from an external cavity. 
As a further development, the linewidth enhancement factor $\alpha$ is included to model the coupling between the amplitude and phase fluctuations of the electric field (which is dependent on the refractive index and optical gain in the laser medium) \cite{masoller1997implications}. Finally, since our system represents the interaction between $M$ coupled lasers, we can now express the LK equations for each laser $j=1,\ldots,M$ as 
\begin{equation}
\begin{aligned}
    \dot{E}_j(t)&=  \frac{1+i \alpha_j}{2}\left(G_j-\gamma \right) E_j(t)+ i\omega_j E_j(t)+ {\kappa_j}\sum_{k=1}^M {A}_{j k} E_k(t-\tau_{jk}),
    \\
    \dot{N}_j(t) &=  J_{0}-\gamma_{n} N_j(t)-G_j\left|E_j(t)\right|^2,
\end{aligned}
\end{equation}
\noindent
where $A =(A_{ij})$ is the adjacency matrix of the coupling network used throughout the paper and $\tau_{jk}$ represents the light travel time from laser $k$ to $j$. By assuming that $\tau_{jk}\approx \tau$ for all pairs $(j,k)$, we obtain the final form of the LK equations, as reported in polar coordinates in Eq. (1).
%

\begin{figure}[t]
\centering
\includegraphics[width=0.55\linewidth]{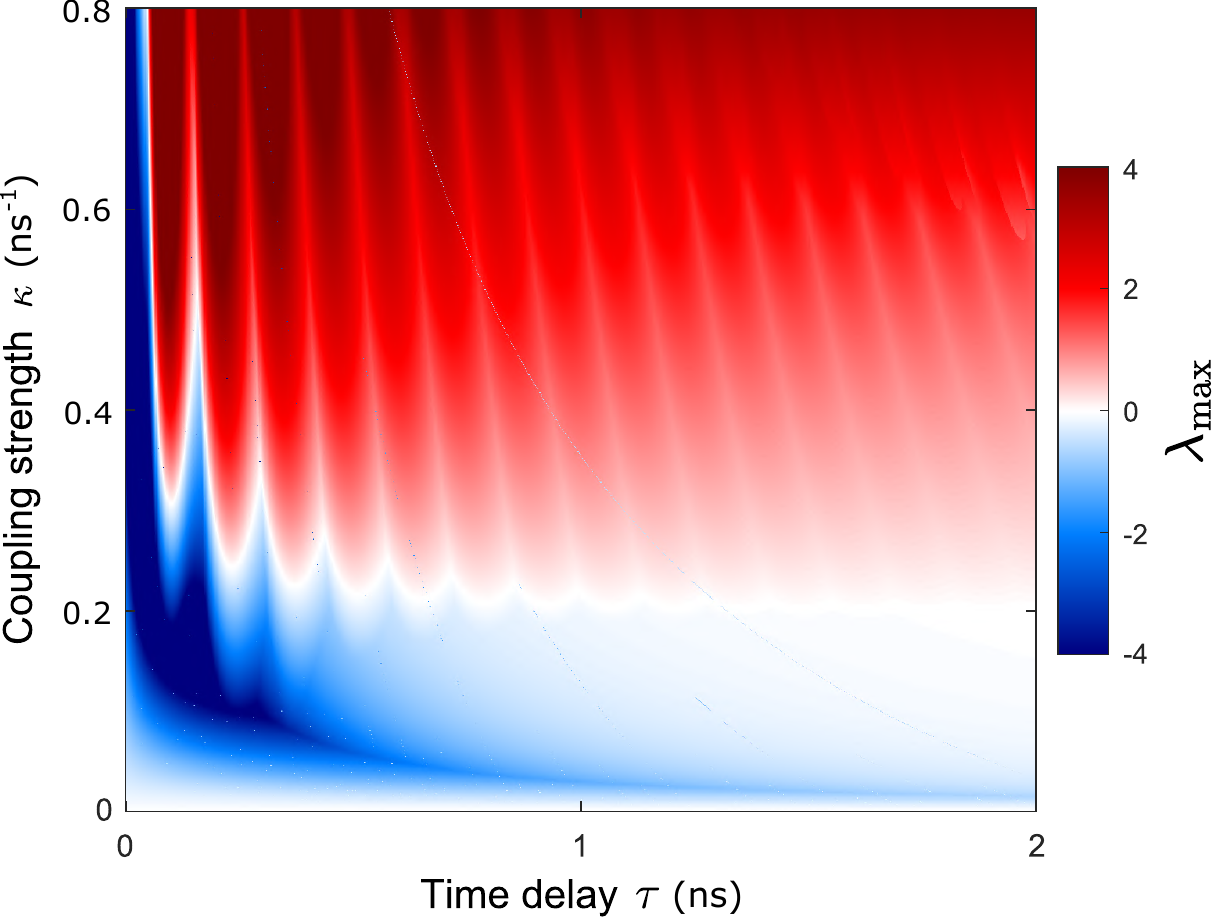}
\caption{Stability of a homogeneous laser array for varying coupling strength and time delay. 
Lyapunov exponent $\lambda_{\rm max}$ of the identical synchronization state (i.e., $E_j(t) = r^* e^{i\Omega_{\rm ML} t}$, $\forall j$) as a function of the coupling strength $\kappa$ and time delay $\tau$, where blue (red) regions correspond to stable (unstable) states. The stability calculation is for the synchronous state corresponding to the minimum linewidth mode $\Omega_{\rm ML}$, considering an all-to-all network with 10 nodes.}
\label{fig:ArnoldTonges}
\end{figure}

\medskip
\textbf{Multistability of the coupled LK model.}
As shown in Fig.~2(a), the LK model admits multiple stationary solutions satisfying $r_j(t)=r^*,$ $N_j(t)=N^*$, and $\phi_j=\Omega t$, which are distributed in an elliptical configuration. 
Fig. \ref{fig:multistability_delay} shows that increasing the time delay or the coupling strength leads to a larger number of solutions, with the elliptical configuration becoming more pronounced. We observe that as $\tau$ increases, the solutions with higher amplitude $r^*$ become increasingly stable. However, the minimum linewidth mode\textemdash the primary focus of our analysis\textemdash consistently appears on the right side of the ellipse and remains unstable over a wide range of parameters.
Fig.~\ref{fig:ArnoldTonges} shows the stability landscape of the coupled LK model as a function of the time delay and coupling strength. Importantly, the stability of this system \textit{decreases} for large coupling. Moreover, for large delay, the coupling interval for which $\lambda_{\rm max}$ is minimized becomes narrower. 
Our results generalize the dynamical analysis of a single self-coupled laser in Ref.~\cite[Sec. 2.3]{kane2005unlocking}, showing that the ellipse-shaped distribution of solutions in the $r\times\Omega$ plane and the periodic instability regions in the $\kappa\times\tau$ plane also emerge in systems with $M>1$ lasers.

\section{Numerical analysis of the Lang-Kobayashi model}

%
\textbf{Solving the transcendental equations.} We rescale the following variables and parameters to improve numerical stability and avoid large discrepancies in the order of magnitude among state variables: ${r}_j\leftarrow r_j \times 10^{-2}$, ${N}_j\leftarrow N_j \times 10^{-8}$, ${N}_0\leftarrow N_0 \times 10^{-8}$, ${g}\leftarrow g \times 10^8$, and ${s}\leftarrow s \times 10^4$.
Moreover, we note that $N^*$ can be determined as a function of the amplitudes $r^*$ as follows: 

\begin{equation}
\begin{aligned}
   N^*=\left(\frac{g r^2}{1+s r^2}+\gamma_n\right)^{-1}\left(J_0+\frac{g r^2}{1+s r^2} N_0\right).
\end{aligned}
   \label{N_in_terms_of_r}
\end{equation}

\noindent
The transcendental equations are thus reduced to $2M$ equations and $2M$ variables (i.e., $r_j^*$ for $j=1,\ldots,M$, $\delta^*_j$ for $j=2,\ldots,M$, and $\Omega$, where $\delta^*_1=0$ without loss of generality). 
We employ MATLAB's nonlinear solver \texttt{fminsearch} to find the set of solutions satisfying Eq. (2) numerically. The stationary solution corresponding to the minimum linewidth mode $\Omega_{\rm ML}$ is determined by solving Eq. (2) over 100 realizations with initial conditions $r_j\sim\mathcal U[0,10]$, $\delta^*_j\sim\mathcal  U[-\pi,\pi]$, and $\Omega\sim\mathcal U[-10,10]$, where $\mathcal U[a,b]$ denotes a uniform distribution within $[a,b]$, and selecting the minimum value $\Omega_{\rm ML} = \min |\Omega|$.

\medskip
\textbf{Numerical continuation and bifurcation analysis.} Introducing disorder to the parameters $\alpha_j$, $\omega_j$, and $\kappa_j$ changes the synchronous solution $E_j(t) = r_j^* e^{i(\Omega_{\rm ML} t + \delta^*_j)}$ according to the transcedental equations (2). In Fig. 3, we employ a \textit{numerical continuation} of the synchronous solution to guarantee that we are tracking the stability of the exact same attractor as the system disorder increases.
This procedure ensures that we stabilize the synchronous state $E_j(t) = r_j^* e^{i(\Omega_{\rm ML} t + \delta^*_j)}$ that is the ``closest'' to the identical state $E_j(t) = r^* e^{i\Omega_{\rm ML} t}$. In this way, both phase and amplitude mismatches among oscillators are minimized.

To formulate the numerical continuation, we express Eq. (2) as a system of parameterized equations
\begin{equation}
\begin{aligned}
    \textbf{G}(\textbf{u}, \sigma_p) = 0
\end{aligned}
   \label{BifAn_TranscEq}
\end{equation}
\noindent where $\textbf{u}$ is the state vector $\textbf{u} = (r^*_1,\hdots,r^*_M,\Omega,\delta^*_1,\hdots,\delta^*_M)^{\top}$ and $\sigma_p$ is a scalar ``control'' parameter (i.e., the level of disorder). We employ the natural parameter continuation method, which is based on iteratively solving the parameterized system \eqref{BifAn_TranscEq}. 
Starting from a system with identical lasers ($\sigma_p^{(0)} = 0$), we solve Eq.~\eqref{BifAn_TranscEq} to obtain the solution $\mathbf u^{(0)}$ corresponding to identical synchronization (where $r_j^*=r^*$ and $\delta^*_j=0$, $\forall j$). Then, we proceed iteratively, for each step $k$, by: i) slightly adjusting the control parameter $\sigma_p^{(k)} \leftarrow \sigma_p^{(k-1)} + \Delta\sigma_p$, ii) solving the corresponding Eq.~\eqref{BifAn_TranscEq}, and iii) finding the (non-identical) solution $\mathbf u^{(k)}$ that is the closest to the previous solution $\mathbf u^{(k-1)}$. The continuation iteratively proceeds over the entire range of $\sigma_p$, although it might fail at points $(\textbf{u},\sigma_p)$ in which the Jacobian matrix $\operatorname{D}_{\textbf{u}}\textbf{G}$ is singular. This indicates that the solution undergoes a fold bifurcation and stops existing from that point forward \cite{dickson2007condition,green2009stability}. Therefore, some of the curves stop abruptly when $\det (\operatorname{D}_{\textbf{u}}\textbf{G})=0$ (as shown in Fig.~\ref{fig:AllCurves}) due to the preclusion of the natural parameter continuation.

\medskip
\textbf{Stability analysis of time-delay systems.} The stability analysis of the synchronous state is based on the variational equation (4), whose Jacobian matrices are given by
\begin{equation}
\begin{aligned}
\operatorname{D}_{\textbf{x}_j}^{(0)}\textbf{f}_j  &= \begin{bmatrix}
\frac{1}{2} \left(G_j\frac{ 1-s r_j^{*2}}{1+s r_j^{*2}}-\gamma \right) & 0 &  \frac{r^*_j}{2}  \frac{G_j}{N_j-N_0}\\
-\alpha \frac{G_j}{1+s r_j^{*2}} sr^*_j & 0 & \frac{\alpha}{2}\frac{G_j}{N_j-N_0}\\
-2r^*_j \frac{G_j}{1+s r_j^{*2}} & 0 &  -\left(\gamma_n+ r_j^{*2} \frac{G_j}{N_j-N_0}\right) \\
\end{bmatrix},
\\
\operatorname{D}_{\textbf{x}_j}^{(0)}\textbf{h}_j &= 
\begin{bmatrix}
0 & \sum_k A_{jk} r^*_k \sin \Phi_{jk} & 0\\
-\sum_k A_{jk} \frac{r^*_k}{r_j^{*2}} \sin \Phi_{jk} & -\sum_k A_{jk} \frac{r^*_k}{r^*_j} \cos \Phi_{jk} & 0\\
0 & 0 & 0 \\
\end{bmatrix},
\\
\operatorname{D}_{\textbf{x}_k}^{(\tau)}\textbf{h}_j &= A_{jk}
\begin{bmatrix}
\cos\Phi_{jk} & -r^*_k\sin\Phi_{jk} & 0\\
\frac{1}{r^*_j}\sin\Phi_{jk} & \frac{r^*_k}{r^*_j}\cos\Phi_{jk} & 0\\
0 & 0 & 0 \\
\end{bmatrix},
\end{aligned}
\end{equation}

\noindent 
where $\Phi_{jk} = \delta^*_k - \delta^*_j - \Omega\tau$. Accordingly, we have that the time-independent matrices in Eq.~(5) are determined element-wise as 
$({J}_{1})_{jk}=  \operatorname{D}_{\textbf{x}_j}^{(0)}\textbf{f}_j\delta_{jk} + \operatorname{D}_{\textbf{x}_j}^{(0)}\textbf{h}_j$
and
$({J}_{2})_{jk}=  \operatorname{D}_{\textbf{x}_k}^{(\tau)}\textbf{h}_j$, where $\delta_{jk}$ is the Kronecker delta.
To numerically calculate the characteristic exponents $\lambda_\ell$ of Eq. (6), we employ the function \texttt{ddebiftool\_stst\_stabil} provided in the MATLAB DDE-BIFTOOL package for the analysis of delay differential equations \cite{engelborghs2000numerical}. 

\medskip
\textbf{Multistability analysis.} To understand how disorder affects \textit{all} stationary solutions satisfying $E_j(t) = r_j^* e^{i(\Omega t + \delta^*_j)}$ (not just the minimum linewidth mode $\Omega_{\rm ML}$), we evaluate the stability of these solutions as $\sigma_{\omega}$ is increased through numerical continuation. Specifically, we focus on the singularity conditions of the Jacobian matrix $\operatorname{D}_{\textbf{u}}\textbf{G}$, measured by its condition number, in order to identify bifurcation points. The condition number of a matrix is defined as $\rm{cond}(A) = \frac{max(eig(A))}{min(eig(A))}$. A smaller inverse condition number indicates proximity to singularity, signaling a breakdown of the numerical continuation.

\begin{figure}[!t]
\centering
\includegraphics[width=0.85\linewidth]{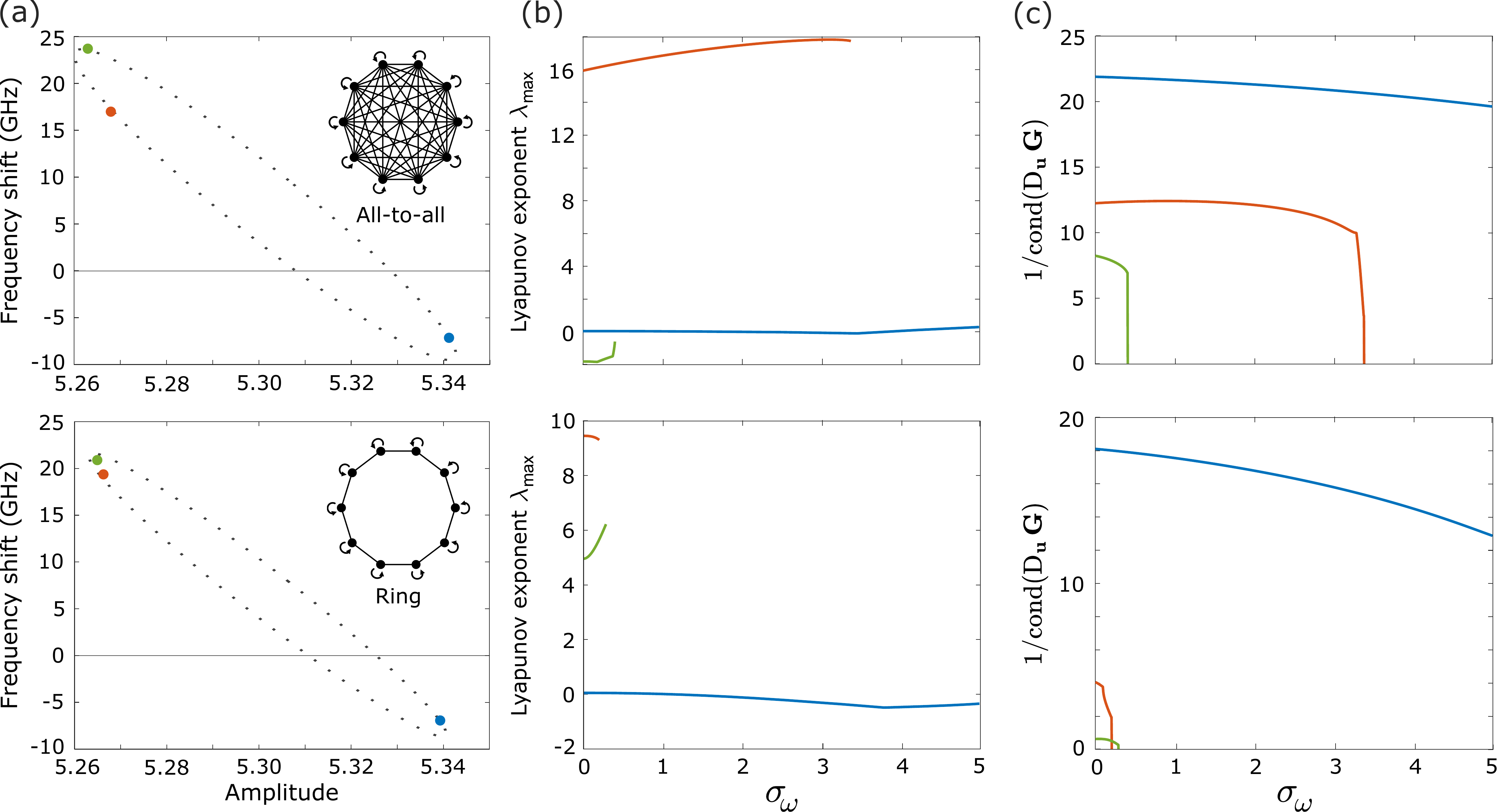}
\caption{Stability and bifurcation analysis for each stationary solution. (a) Spectrum of stationary solutions (amplitude–frequency pairs) for all-to-all and ring topologies with 10 lasers. \rev{The parameters are set as $\tau = 0.15$ ns, $\kappa = 0.47\, \rm{ns}^{-1}$ (all-to-all), and $\kappa = 1.44\, \rm{ns}^{-1}$ (ring).} 
(b)~Corresponding Lyapunov exponents for each stationary solution under the same realization of frequency detuning disorder $\omega_j$.
(c)~Inverse of condition number of $\operatorname{D}_{\textbf{u}}\textbf{G}$.
Each of the stationary solutions is color-coded by red, green, or blue across all panels.}
\label{fig:AllSol}
\end{figure}

\begin{figure}[!t]
\centering
\includegraphics[width=0.7\linewidth]{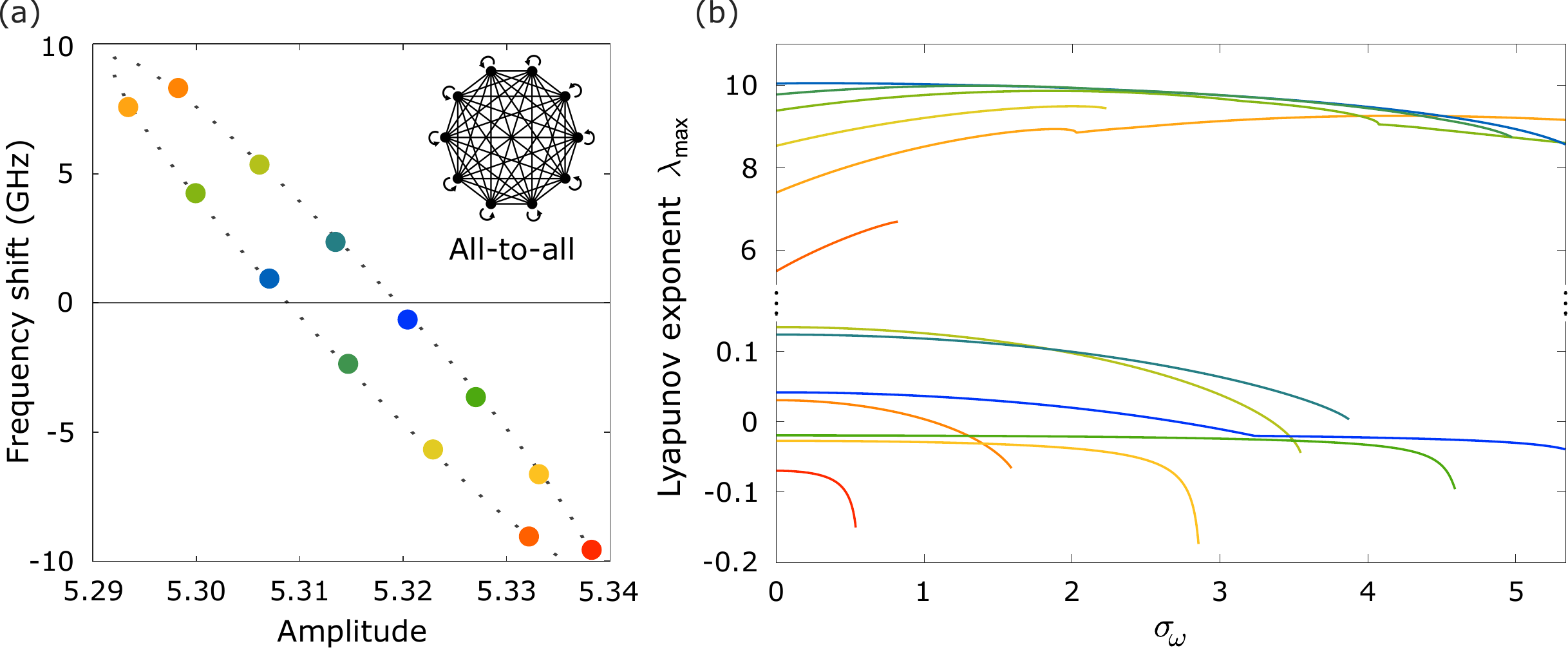}
\rev{\caption{Stability and bifurcation analysis in a highly multistable regime.
(a) Spectrum of stationary solutions (amplitude–frequency pairs) for an all-to-all topology with 10 lasers. The parameters are set as $\tau = 2$ ns and $\kappa = 0.21\, \rm{ns}^{-1}$. 
(b)~Corresponding Lyapunov exponents for each stationary solution under the same realization of frequency detuning disorder $\omega_j$. The top curves correspond to the solutions on the left side of the ellipse, which exhibit significantly higher instability, and the bottom curves represent the solutions on the right side, which are closer to the stability threshold.
The curve colors correspond to the solutions in the left panel, gradually changing from red to blue as the frequency approaches zero.}
\vspace{-0.4 cm}
\label{fig:AllSolHet}}
\end{figure}

\rev{We first consider a parameter regime with small $\kappa$ and $\tau$, so that the number of coexisting equilibria is small.} For both all-to-all and ring networks, Fig.~\ref{fig:AllSol} evaluates the largest Lyapunov exponent and the singularity of $\operatorname{D}_{\textbf{u}}\textbf{G}$ for each of the 3 stationary solutions (including the minimum linewidth mode) as disorder is increased. \rev{Notably, all solutions other than the minimum linewidth mode undergo bifurcation and vanish at lower levels of disorder.} In the all-to-all configuration, the solution with intermediary $\Omega$ (orange curve) is unstable and does not benefit from heterogeneity, whereas the solution with the largest $\Omega$ (green curve) starts as a stable solution but soon ceases to exist at a very low level of disorder. In the ring configuration, both of the solutions vanish at very small levels of disorder. 
\rev{For systems with many coexisting equilibria (high $\kappa$ and $\tau$), Fig. \ref{fig:AllSolHet} also shows that the branch originating from the minimum-linewidth mode is the only one that simultaneously transitions to stability and persists under higher levels of disorder. All other solutions are either intrinsically unstable or sufficiently shifted from the ideal frequency, vanishing at lower heterogeneity levels. Solutions on the left side of the ellipse can persist under higher levels of heterogeneity but are intrinsically more unstable. In contrast, solutions on the right side lie closer to the stability threshold and consistently benefit from disorder, although they undergo bifurcation earlier than the minimum-linewidth mode.
This bifurcation analysis suggests that disorder can be leveraged to systematically stabilize the minimum-linewidth mode over a broad range of multistable regimes.}

\section{Synchronization analysis and irregular networks}
\label{sec.ParamHet}

\begin{figure}[t]
\centering
\includegraphics[width=0.75\linewidth]{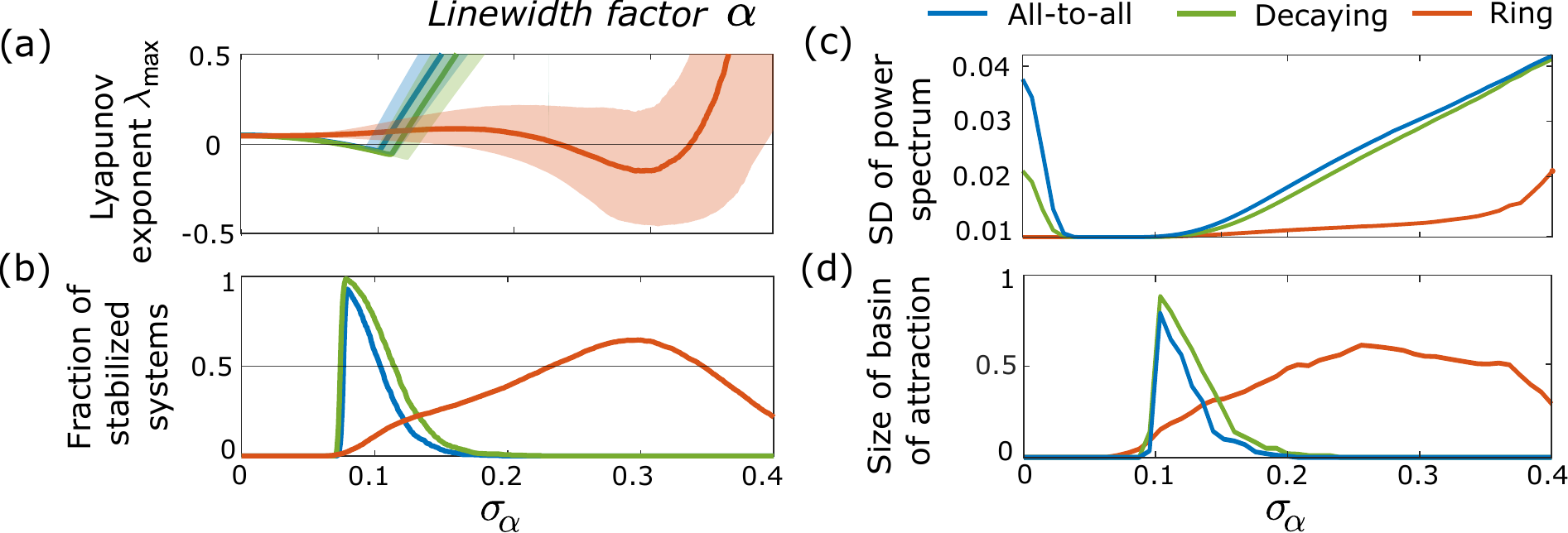}
\caption{Synchronization stability for varying levels of disorder in the linewidth factor.
(a) Lyapunov exponent $\lambda_{\rm max}$ as a function of the level of disorder $\sigma_p$ introduced to the linewidth factor $\alpha$. The results are shown for three network topologies with $M=10$ lasers: all-to-all (blue), decaying (green), and ring (orange). The solid lines indicate the median values across 1,000 independent realizations of parameter disorder, while shaded areas indicate the first and third quartiles. 
(b) Fraction of realizations that successfully stabilize the frequency-synchronized state (for which $\lambda_{\rm max}(\sigma) < 0$). 
(c)  Average standard deviation of the power spectrum of the combined field $E(t)$ (in steady state) for random initial conditions.
(d) Average size of the basin of attraction.}
\label{fig:Alpha_het}
\end{figure}

\begin{figure}[t]
\centering
\includegraphics[width=0.8\linewidth]{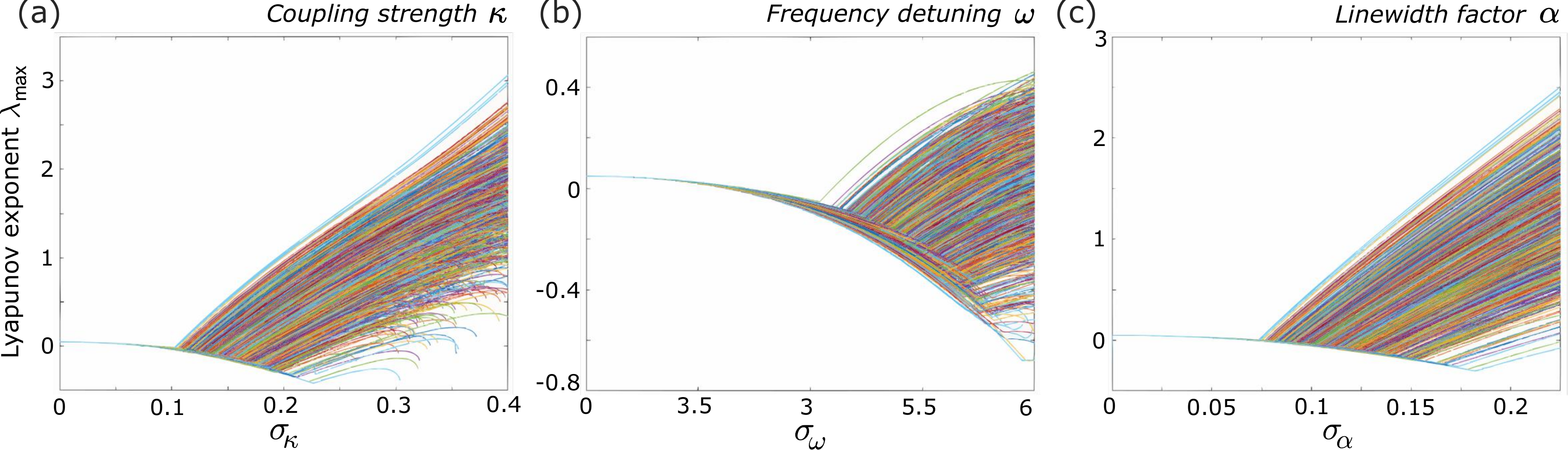}
\caption{Synchronization stability across individual realizations for varying levels of disorder. Lyapunov exponent $\lambda_{\rm max}$ as a function of $\sigma_p$ for each of the 1,000 realizations in a decaying network of 10 lasers. In each panel, disorder is introduced to (a) the coupling strength, (b) the frequency detuning, and (c) the linewidth factor.
Note that some curves terminate earlier; such realizations correspond to cases in which the numerical continuation of the solution $E_j=r_j^* e^{i(\Omega_{\rm ML}t + \delta^*_j)}$ leads to a bifurcation in which it ceases to exist.}
\vspace{-0.4 cm}
\label{fig:AllCurves}
\end{figure}

\subsection{Unconstrained disorder}
\label{sec.AllTrials}

In this section, we provide numerical details for the simulations presented in Figs.~3 and 4, as well as an analysis of introducing disorder to the linewidth enhancement factor $\alpha$ (Fig.~\ref{fig:Alpha_het}).  The parameter disorder is set as follows in each scenario: $\alpha_j\sim\mathcal N(5,\sigma^2_\alpha)$, $\omega_j\sim\mathcal N(0,\sigma^2_\omega)$, and $\kappa_j\sim\mathcal N(\kappa_{\rm hom},\sigma^2_\kappa)$. The homogeneous coupling $\kappa_{\rm hom}$ is chosen to be sufficiently large for each network topology so that the largest Lyapunov exponent is slightly positive ($\lambda_{\rm max}\approx 0.05$):  $\kappa_{\rm hom}=0.475\, \rm{ns}^{-1}$ for the all-to-all topology, $\kappa_{\rm hom}=0.538\, \rm{ns}^{-1}$ for the decaying topology, and $\kappa_{\rm hom}=1.44\, \rm{ns}^{-1}$ for the ring topology. We set $\tau = 0.15$~ns, while other constructive parameters of the LK model are reported in the main text.
Fig. \ref{fig:AllCurves} individually shows the Lyapunov exponent as a function of the level of disorder $\sigma_p$ for each of the 1,000 disorder realizations, with the median values displayed in Figs.~3(a) and \ref{fig:Alpha_het}(a).
Examining these curves across all realizations, we find that the main variation lies in the specific level of disorder at which stability reaches its optimal point, as well as the bifurcation point at which the corresponding solutions cease to exist.

\begin{figure}[t]
\centering
\includegraphics[width=0.45\linewidth]{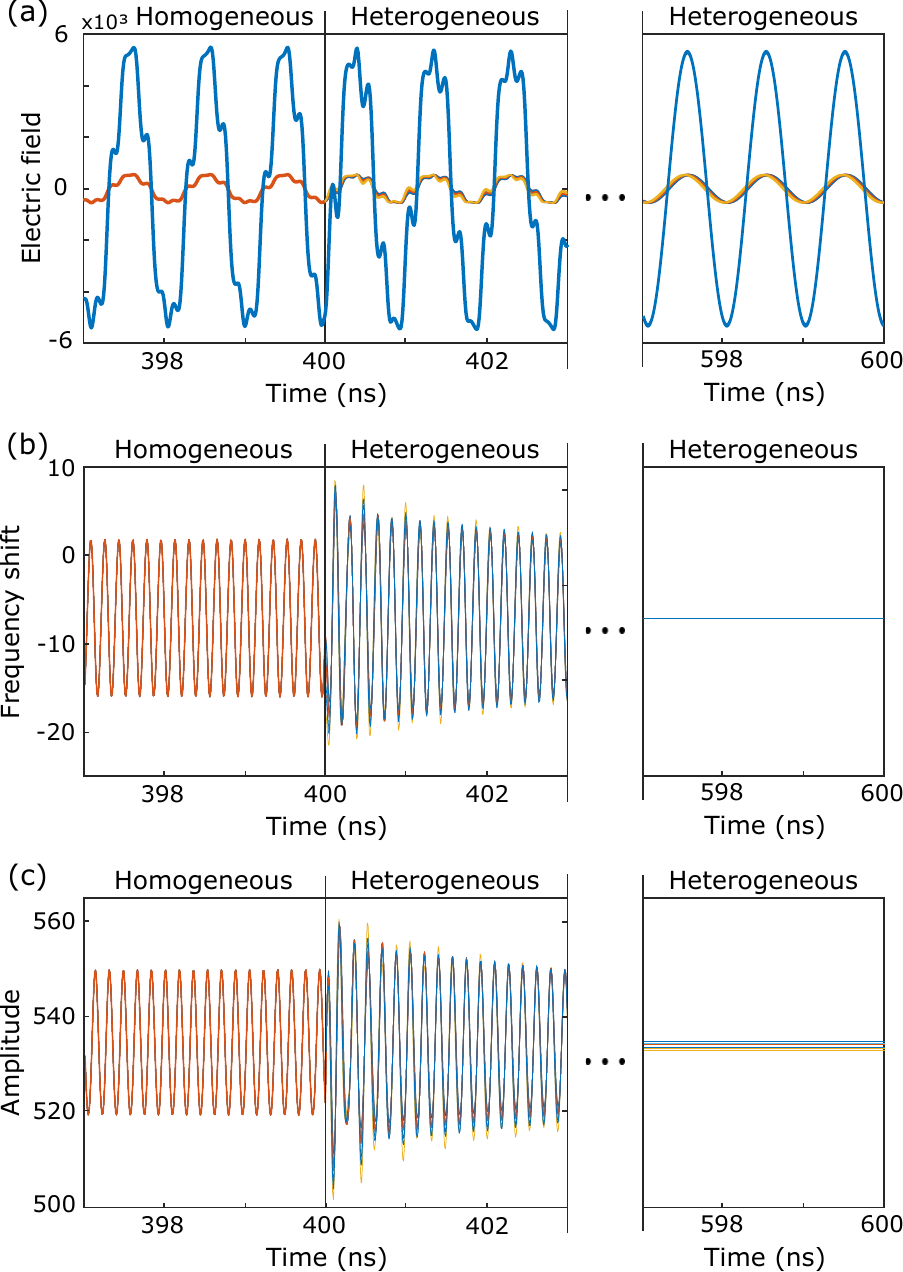}
\rev{\caption{Time series of homogeneous vs heterogeneous lasers. Time series of the (a) electric fields, (b) frequency shifts, and (c) amplitudes of the coupled lasers transitioning from a homogeneous configuration for $t\in [0,400]$  ns (in which $\omega_j=0,\forall j$) to a heterogeneous configuration for $t\in[400,600]$ ns (in which $\omega_j$ follows a normal distribution). These plots expand on the results presented in Fig.~2(b).}
\vspace{-0.4 cm}
\label{fig:TimeSeries}}
\end{figure}

\medskip
\rev{\textbf{Time-series simulation.}
To complement the dynamical simulations shown in Fig.~2(b) and highlight the transition from the homogeneous to the heterogeneous parameter configuration, we also present in Fig. \ref{fig:TimeSeries} the time series of the instantaneous frequency and amplitude of the lasers over the same time interval. In Fig. \ref{fig:TimeSeries}(a), we observe that, immediately after introducing heterogeneity, the lasers enter a transient regime where their signals initially desynchronize and become less regular. However, as the system relaxes toward a stationary state, the electric fields evolve into smoother, sinusoidal-like waveforms: the frequencies lock to a common constant value across all lasers (as the relative phase differences stay fixed), and the amplitudes stabilize. This contrast is clearer in Figs.~\ref{fig:TimeSeries}(b)-(c): initially, both frequency and amplitude exhibit large oscillations, but after the transient, these oscillations vanish.}

\medskip
\textbf{Power spectrum.}
To calculate the power spectrum presented in Figs. 3(c) and \ref{fig:Alpha_het}(c), we first simulate the system dynamics over the time interval $t\in[0, 300]$~ns using the MATLAB solver \texttt{dde23} (for time-delay differential equations). The initial conditions are randomly drawn around the synchronous state with small perturbations of magnitude $10^{-5}$, i.e., $r_j\sim\mathcal N(r_j^*,10^{-10})$, $N_j\sim\mathcal N(N_j^*, 10^{-10})$, $\delta_j\sim\mathcal N(\delta_j^*, 10^{-10})$, and $\Omega\sim\mathcal N(\Omega_{\rm ML}, 10^{-10})$. 
We then obtain the combined electric field by summing over all the lasers and calculate the corresponding power spectrum using the \texttt{pspectrum} function from MATLAB's Signal Processing toolbox. The standard deviation of the power spectrum is given by $\sigma_{\rm PS} = \sqrt{\sum_j{P_jf_j^2 }}$, where $P_j$ is the normalized power associated with the frequency value $f_j$. 

We also use the power spectrum analysis to obtain the results presented in Fig. 4. For large-scale laser networks comprising up to a thousand lasers, it is computationally unfeasible to estimate the Lyapunov exponent $\lambda_{\rm max}$. Thus, as a proxy measure of stability, we simulate the system dynamics (with initial conditions close to the synchronous state) and compute the power spectrum associated with the stationary solution after a simulation time of 300 ns. We consider a system to be synchronized if $\sigma_{\rm PS}\leq 0.01$, which indicates that the stationary solution is dominated by a single-mode frequency.


\medskip
\textbf{Attraction basins.}
For non-delay systems, the basin of attraction can be numerically estimated by simulating the system dynamics over a wide range of initial conditions and verifying which of these initial conditions converge to (or diverge from) the desired stationary solution $\textbf{x}^*$. 
However, for time-delay systems, the state space has infinite dimensions since the initial conditions are determined by the entire ``history'' of the state variables within the interval $t\in[-\tau, 0)$ \cite{leng2016basin,rakshit2017basin}.
We investigate the convergence of the system dynamics to the frequency-synchronized state by randomly selecting initial conditions with amplitudes $r_j(0)\sim\mathcal U[520, 540]$, frequency shifts $\Omega_j(0)\sim\mathcal U[-10, 30]$ GHz, and phases $\delta_j(0) \sim \mathcal U[-\pi,\pi]$, for $j=1,\ldots,M$, where $\mathcal U[a,b]$ denotes a uniform distribution within intervals $[a,b]$. These uniform intervals were chosen to cover the entire range of frequency-synchronized solutions depicted by the ellipse in Fig. 2(a). Thus, the initial condition of the system is given by $E_j(t) = r_j(0) e^{i(\Omega_j(0)t + \delta_j(0))}$ for $t\in[-\tau, 0)$, whereas $N_j(0)$ is determined by Eq.~\eqref{N_in_terms_of_r} in steady state. The basin of attraction is quantified as the fraction of 200 randomly generated initial conditions that converge to the synchronous state, as shown in Figs. 3(d) and \ref{fig:Alpha_het}(d). 

\begin{figure}[t]
\centering
\includegraphics[width=0.85\linewidth]{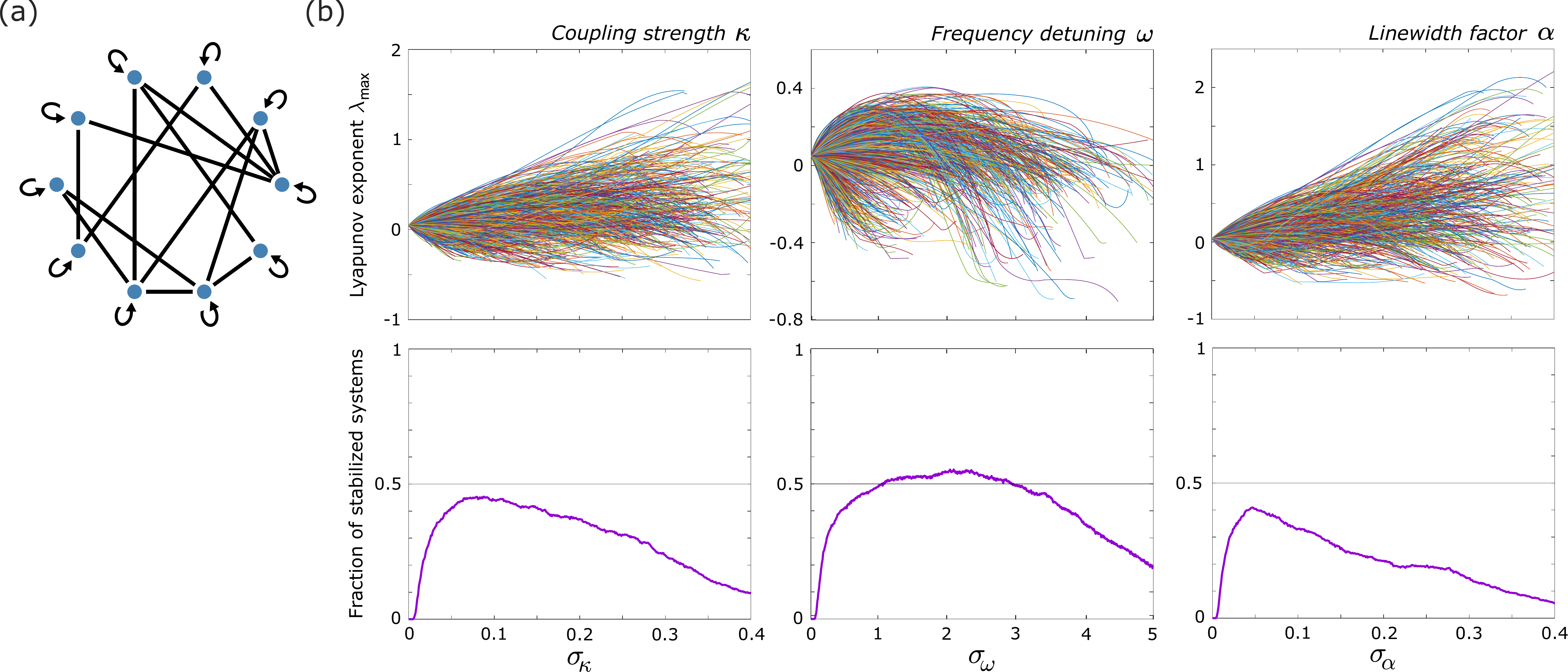}
\caption{Synchronization stability promoted by disorder for an arbitrary network structure.  (a) Random network in which two nodes are connected with probability $p=0.3$. (b) Lyapunov exponent $\lambda_{\rm max}$ (top) and fraction of stabilized systems (bottom) as functions of $\sigma_p$ across 1,000 realizations. In each panel, disorder is introduced to the coupling strength (left), frequency detuning (center), and linewidth factor (right).
}
\label{fig:ArbNet}
\end{figure}

\begin{figure}[!t]
\centering
\includegraphics[width=0.7\linewidth]{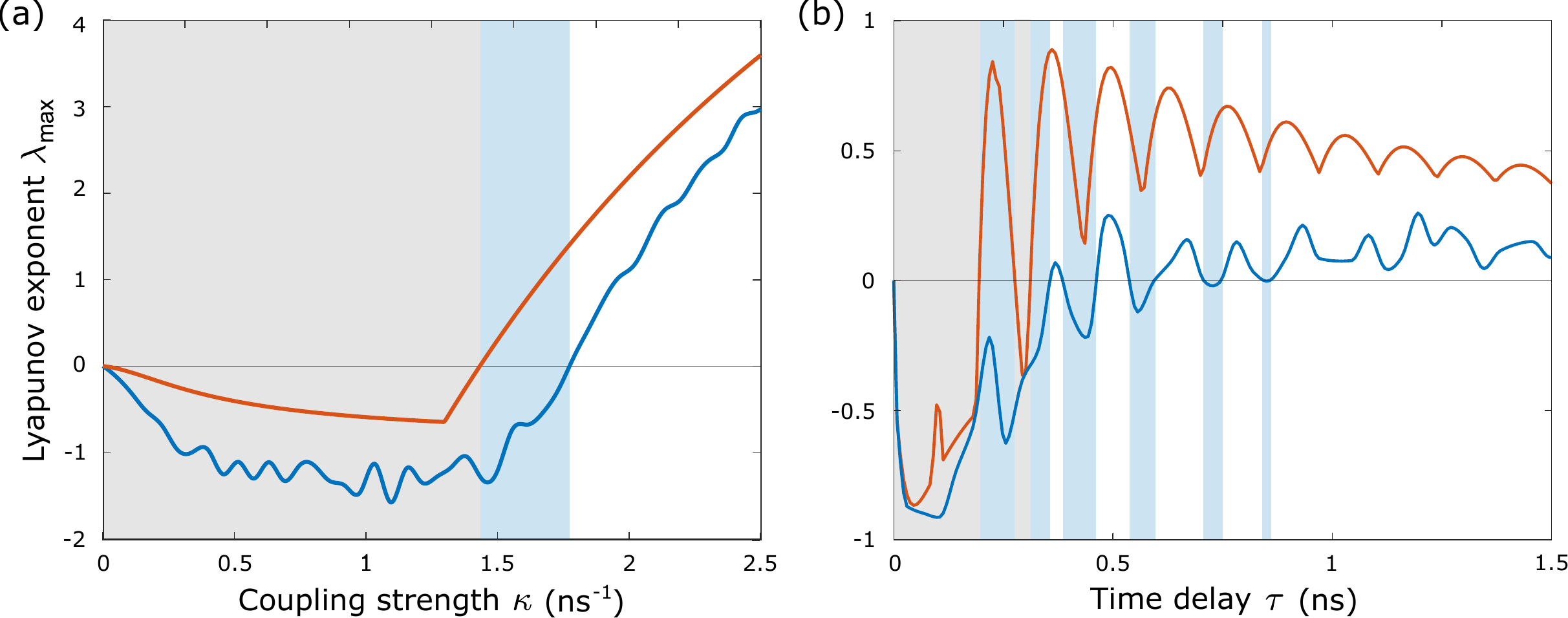}
\rev{\caption{Stability regions as a function of coupling strength and time delay for homogeneous and heterogeneous systems. Lyapunov exponent of the synchronized state for homogeneous (orange) and heterogeneous (blue) parameter configurations as a function of (a) coupling strength $\kappa$ (with $\tau = 0.15$ ns) and (b) time delay $\tau$ (with $\kappa = 1.1\, \rm{ns}^{-1}$). The results are shown for a 10-node ring network. For each value of $\kappa$ or $\tau$, the plotted curve corresponds to the realization that achieves maximum stability out of 100 random realizations of frequency detuning. The shaded areas in gray represent the regions where both the homogeneous and heterogeneous systems are stable, and the areas in blue represent the regions where only the heterogeneous systems are stable.}
\vspace{-0.5 cm}
\label{fig:KappaRange}}
\end{figure}

\medskip
\textbf{Irregular network.} We also extend our analysis to a representative (Erdős–Rényi) random network, generated by adding undirected edges between nodes with probability $p=0.3$ and setting the self-coupling as $A_{ii}=1$. Fig. \ref{fig:ArbNet} shows the Lyapunov exponent as a function of the disorder level $\sigma_p$ for each of the 1,000 realizations. Unlike the results in Fig.  \ref{fig:AllCurves}, $\lambda_{\rm max}$ does not immediately decrease as a function of $\sigma_p$, and the transition to a synchronous state may occur after the curve reaches a local maximum. Furthermore, the stability curves do not exhibit a consistent trend due to the loss of network symmetry (compared to the regular network structures studied in Fig. 3). As we discuss in Sec. \ref{SEC:LyapSurfPlots}, when the network is symmetric (Fig.  \ref{fig:LyapSurf}), for any parameter realization $\delta p$, there exists a disorder level $\sigma$ for which the system achieves stability; however, this is not necessarily the case for asymmetric networks (Fig. \ref{fig:LyapSurfTopo}).

\medskip
\textbf{Stability range of coupling strength \rev{and time delay}.} \rev{Figure~\ref{fig:KappaRange} shows that the stable synchronous state can be obtained over a wider range of coupling strengths and time delays for disordered laser systems compared to their homogeneous counterparts.} For homogeneous systems, the largest Lyapunov exponent associated with the identical synchronous solution changes as a function of the coupling strength, becoming unstable at \rev{$\kappa = 1.44\, \rm{ns}^{-1}$}. Introducing disorder in the frequency detuning $\omega$\textemdash considering the best-case realization\textemdash extends the critical threshold to \rev{$\kappa= 1.77\, \rm{ns}^{-1}$}, thereby increasing the stability region by 23.7\%. \rev{Likewise, Fig.~\ref{fig:KappaRange}(b) shows that over a wide range of time delays, heterogeneity yields significant stability improvement, extending the range of stability from $\tau = 0.19$ ns to $\tau = 0.36$ (a relative improvement of 89.4\%).}

\subsection{Constrained disorder} 
Here, we consider two different strategies for introducing parameter disorder and assess their impact on stability. The first approach focuses on unconstrained random parameter perturbations as reported above and presented in Figs. 3 and \ref{fig:Alpha_het}. The second approach is to constrain the parameter perturbations such that the solution of the transcendental equation corresponding to the identical synchronization $E_j = r^* e^{i\Omega_{\rm ML}t}$ is preserved. This design ensures that a coupled system of non-identical lasers can still achieve identical synchronization. From the LK model (1), the transformations $\alpha_j \rightarrow \alpha +h_j$ and  $\omega_j \rightarrow \omega +\frac{1}{2\sigma}\left(\gamma - g\frac{N-N_0}{1+sr^2}\right)h_j$ leave the solution invariant for any $h_j\in \mathbb{R}$. 
To implement this constrained approach, we start from a homogeneous system (as in Fig. 3) and introduce heterogeneity by sampling the parameter $h_j\sim\mathcal N(0,\sigma_h)$ for each laser $j$. Importantly, this approach requires simultaneous perturbation of at least two parameters in order to leave the identical solution invariant. 

\begin{figure}[!t]
\centering
\includegraphics[width=0.85\linewidth]{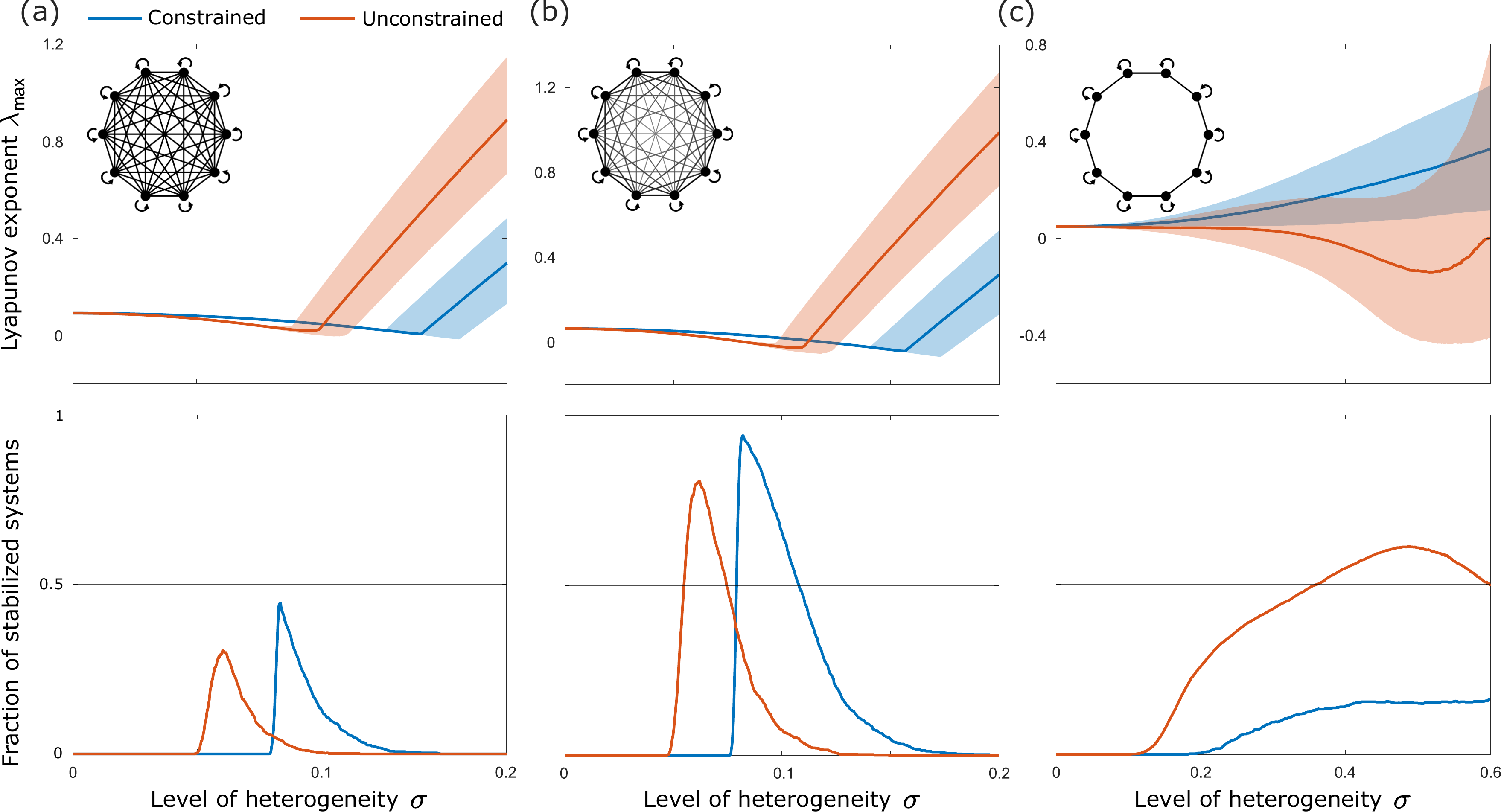}
\caption{Comparison between unconstrained and constrained disorder as mechanisms for synchronization stability.
Lyapunov exponent $\lambda_{\rm max}$ (top) and fraction of stabilized systems (bottom) as functions of the level of disorder $\sigma_d$ for (a) all-to-all, (b) decaying, and (c) ring networks. 
The results compare two approaches to introduce disorder: i) unconstrained perturbations (orange), in which parameters $\alpha_j\sim\mathcal N(5,\sigma^2)$ and $\omega_j\sim\mathcal N(0,\sigma^2)$ are simultaneously and independently perturbed, and ii) constrained perturbations (blue), in which parameters $\alpha_j$ and $\omega_j$ are perturbed by a factor $h_j\sim\mathcal N(0,\sigma^2)$ that leaves the identical synchronization solution invariant. 
The solid lines in the top panels indicate the median across 1,000 realizations, while shaded areas indicate the first and third quartiles.} 
\label{fig:DesignedRandom}
\end{figure}

Fig.~\ref{fig:DesignedRandom} compares the improvement in the synchronization stability when disorder is simultaneously introduced to two parameters (the linewidth factor $\alpha$ and the frequency detuning $\omega$). We compare the unconstrained and constrained strategies. For the all-to-all and decaying topologies, our results show that the stability improves as a function of the disorder level $\sigma$ in both strategies, although only a small fraction of realizations transition from the unstable to the stable regime. On the other hand, for ring topologies, the constrained disorder is detrimental to synchronization stability, as it causes $\lambda_{\rm max}$ to increase on average. Furthermore, a comparison between Figs. 3 and \ref{fig:DesignedRandom} reveals that applying random perturbations to \textit{single} parameters (Fig. 3), rather than \textit{pairs} of parameters (Fig.~\ref{fig:DesignedRandom}), substantially improves the fraction of stabilized systems across a given range of disorder levels $\sigma_p$ (specifically, in Fig. \ref{fig:DesignedRandom}, both perturbed parameters are assigned the same level of heterogeneity).   These results show that unconstrained disorder is a more robust approach to designing coupled laser arrays.

\begin{figure}[t]
\centering
\includegraphics[width=0.82\linewidth]{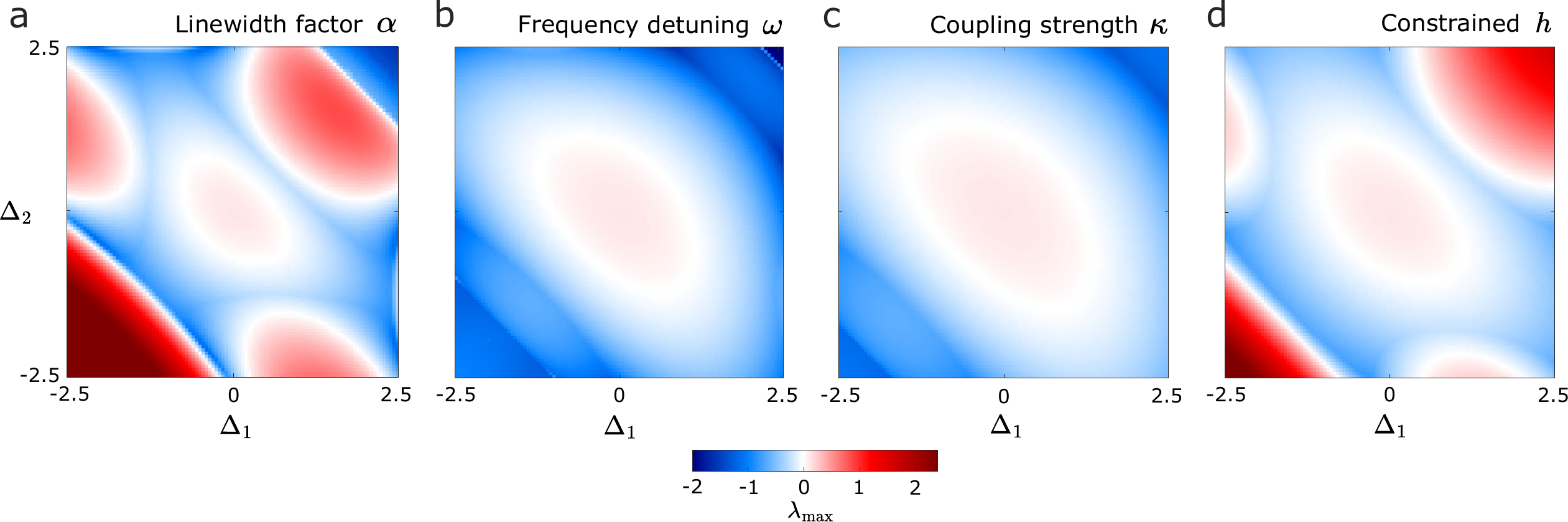}
\caption{Stability landscape for a network of $M=3$ lasers. Lyapunov exponent $\lambda_{\rm max}$ (color coded) as a function of the heterogeneity factors $\Delta_j$ applied to parameters: (\textbf{a}) $\alpha_j$, (\textbf{b}) $\omega_j$, (\textbf{c}) $\kappa_j$, and (\textbf{d}) $h_j$. Panels a--c correspond to unconstrained disorder applied independently to each of the parameters, while panel d corresponds to the constrained disorder $h_j = \Delta_j$ applied simultaneously to parameters $\alpha$ and $\omega$.
}
\label{fig:LyapSurf}
\end{figure}

\begin{figure}[!t]
\centering
\includegraphics[width=0.8\linewidth]{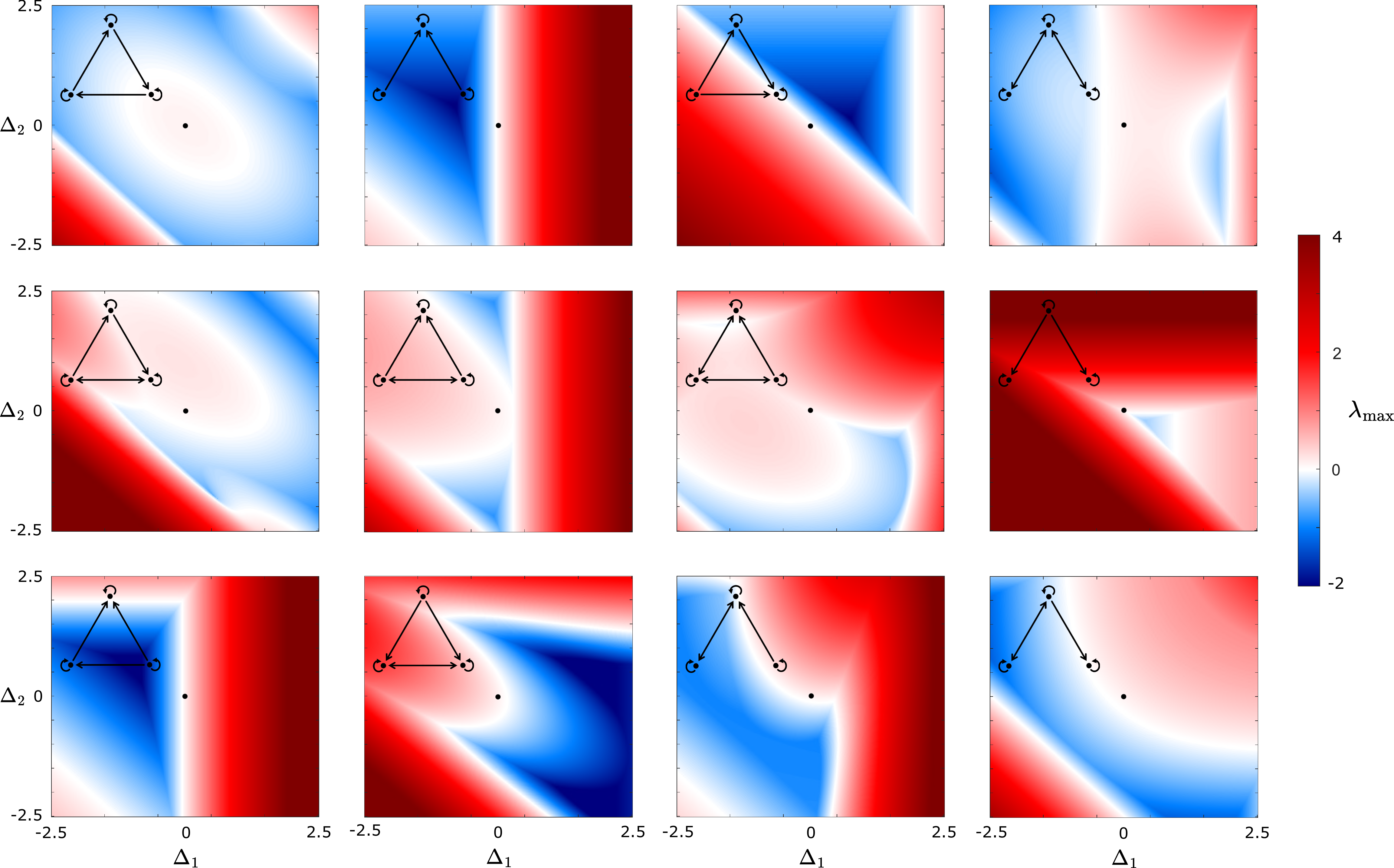}
\caption{Stability landscape for all possible network topologies comprising $M=3$ lasers. Lyapunov exponent $\lambda_{\rm max}$ (color coded) as a function of the constrained heterogeneity factors $h_j = \Delta_j$ applied simultaneously to parameters $\alpha$ and $\omega$. The networks are indicated in the insets. The black dot marks at the origin correspond to the unstable homogeneous system.}
\vspace{-0.4 cm}
\label{fig:LyapSurfTopo}
\end{figure}

\section{Stability landscape in small networks}
\label{SEC:LyapSurfPlots}

For a network of $M=3$ lasers, we analyze the stability of the system when the lasers are nonidentical and globally coupled. 
The laser parameters are set as $p_j = p_{\rm hom} + \Delta_j$, where $p\in\{\alpha,\omega,\kappa\}$ and $\Delta_j$ is a perturbation factor subject to the constraint $\sum_{j=1}^3\Delta_j =0$. This approach allows us to systematically explore the stability of the synchronous state over all possible configurations of heterogeneous parameters by covering the  $\Delta_1 - \Delta_2$ plane, as illustrated in Fig.~\ref{fig:LyapSurf}.
The origin represents the point corresponding to a homogeneous system. As one moves radially outward from the origin, the heterogeneity among the oscillators increases. The highest stability (indicated by the blue regions) occurs far from the origin. In addition, for any combination of parameter heterogeneity $\bm\Delta = \sigma(\Delta_1, \Delta_2, \Delta_3)^\top$ (representing a radial direction from the center), there is a disorder level $\sigma$ such that $\bm\Delta$ lies in a region of stability. However, a small $\sigma$ is insufficient to shift the system from an unstable to a stable state. Conversely, very high $\sigma$ can also destabilize the system. This highlights a balance between synchronizability and heterogeneity, suggesting that intermediate levels of heterogeneity provide robust, stable synchronization with high probability. Such analysis is consistent with the results observed in Fig. 3, in which disorder is randomly added to the parameters as $p_j = p_{\rm hom} + \delta p_j$, where $\delta p_j\sim\mathcal N(0,\sigma_p^2)$, and the system is shown to stabilize for intermediate levels of heterogeneity. Fig.~\ref{fig:LyapSurfTopo} further explores the stability over this parameter space for all possible directed network topologies of $M=3$ lasers, illustrating that the broken symmetries of the network topology lead to asymmetric stability regions.

\section{Effect of disorder in delayed vs. non-delayed class B lasers}
\label{sec.LEfortau0}

In Ref.~\cite{pando2024synchronization}, the authors investigate the synchronization of disordered coupled lasers and conclude that random, uncorrelated disorder among the lasers' parameters (specifically, the frequency detuning $\omega_j$) hinders synchronizability. Given that their theoretical analysis is based on laser rate equations (LRE) without time-delay coupling \cite{rogister2004Power}, this raises an important question in the context of our work: what role do time delay and laser model play in the synchronization stability of disordered lasers? We address this question by comparing the LK and LRE models \rev{for class B lasers}, with and without delay.

\begin{figure}[t]
\centering
\includegraphics[width=0.65\linewidth]{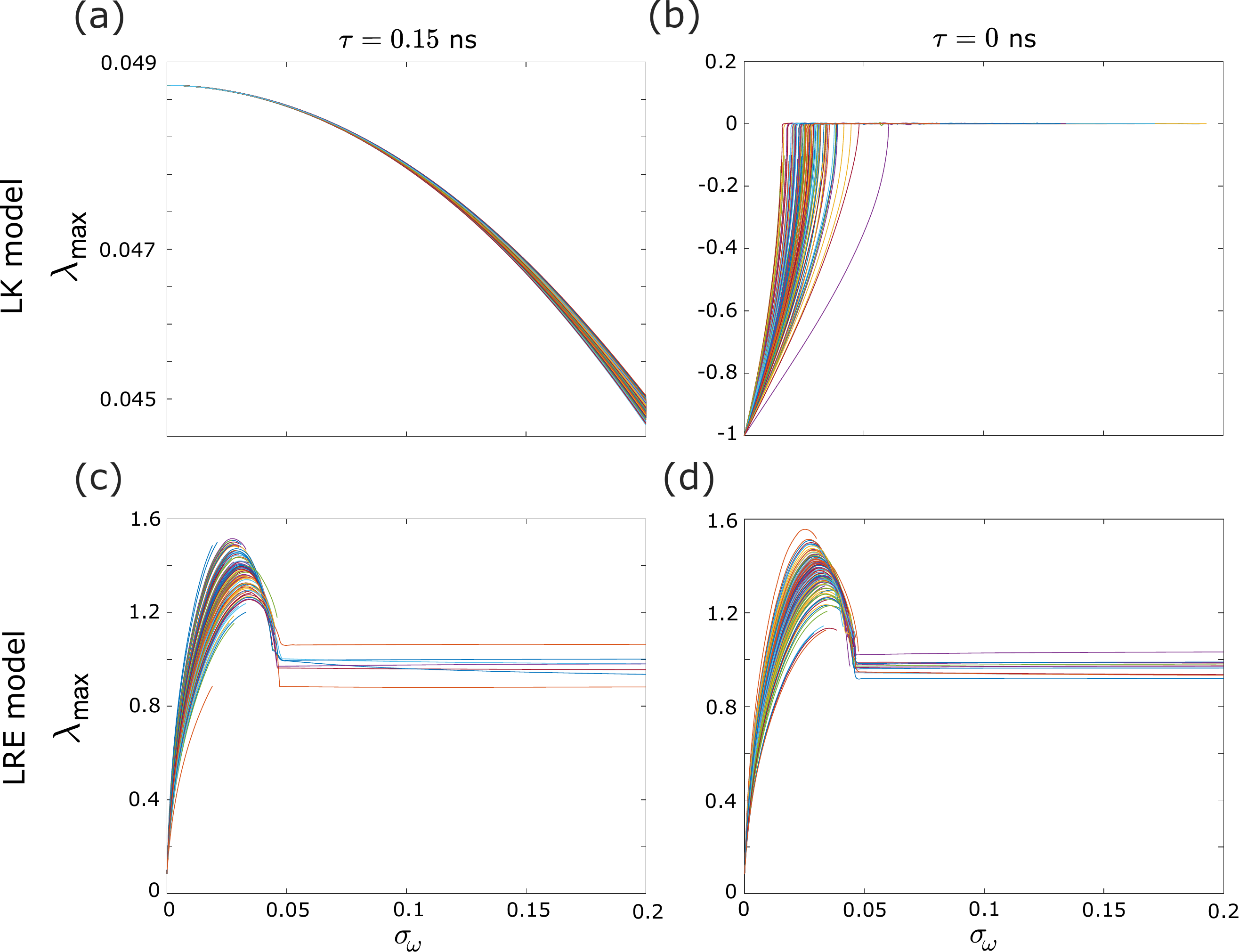}
\caption{Impact of parameter disorder in delayed and non-delayed laser models. Lyapunov exponent $\lambda_{\rm max}$ as a function of $\sigma_\omega$ for each of the 100 realizations in a decaying network with 10 lasers. The top panels show the stability curves for the LK model in the presence and absence of time delay, while the bottom panels show the stability curves for the LRE model.
The parameter values of the LK model are reported in the main text, whereas the parameters of the LRE are set to $J_0 = 19.8$, $\gamma = 1.7$, $s = 0.25$, and $\kappa =0.13\, \rm{ns}^{-1}$.
}
\label{fig:LKvsLREtau0}
\end{figure}

\medskip
\textbf{Non-delayed Lang-Kobayashi model and phase reduction.} 
For the LK model (1), Fig.~\ref{fig:LKvsLREtau0}(a,b) shows opposite trends for the synchronization stability depending on the presence ($\tau = 0.15$) or absence ($\tau = 0$) of time delay. For $\tau = 0.15$, $\lambda_{\rm max}$ decreases as a function of $\sigma_\omega$ , eventually turning negative as shown in Fig.~\ref{fig:AllCurves}(c) for $\sigma_\omega > 0.2$. Conversely, for $\tau = 0$, $\lambda_{\rm max}$ increases, transitioning from a stable homogeneous state ($\lambda_{\rm max}<0$ for $\sigma_{\omega} = 0$) to a marginally stable state ($\lambda_{\rm max}\approx 0$) at large $\sigma_{\omega}$. As we show next, the dynamical behavior of the non-delayed LK model is similar to that of coupled Kuramoto oscillators: the synchronous state is marginally stable when the heterogeneity among oscillators is sufficiently large.

Consider a pair of coupled Kuramoto oscillators:
\begin{equation}
        \begin{aligned}
            \dot\phi_1 &= \omega_1 + K\sin(\phi_2 - \phi_1), \\
            \dot\phi_2 &= \omega_2 + K\sin(\phi_1 - \phi_2),
        \end{aligned}
    \end{equation}
\noindent
where $\phi_i\in\mathbb S^1$ is the phase of oscillator $i$, $\omega_i$ is the oscillator's natural frequency, and $K$ is the coupling strength. The phase difference dynamics follow: $\dot\theta = \frac{d}{dt}(\phi_1-\phi_2) = \Delta\omega - 2K\sin(\theta)$, where $\Delta\omega = \omega_1 - \omega_2 > 0$. Linearizing around equilibrium $\theta^* = \sin^{-1}(\Delta\omega/2K)$, we obtain the variational equation for small perturbations $\delta\theta$ around $\theta^*$:
\begin{equation}
        \delta{\dot\theta} = -2K \cos ( \theta^*).
\end{equation}

\noindent
As the oscillator heterogeneity $\Delta\omega$ increases, $\theta^* \rightarrow \frac{\pi}{2}$ and, therefore, $\delta\dot\theta \rightarrow 0$. Thus, the synchronous state $\theta^*$ becomes marginally stable for sufficiently large $\Delta\omega$. Moreover, when $\Delta\omega > 2K$, the equilibrium point $\theta^*$ reaches a bifurcation point in which it ceases to exist. Such behavior is observed in the non-delayed LK model (Fig.~\ref{fig:LKvsLREtau0}(b)), where many curves terminate earlier for large $\sigma_\omega$.

\begin{figure}[t]
\centering
\includegraphics[width=0.6\linewidth]{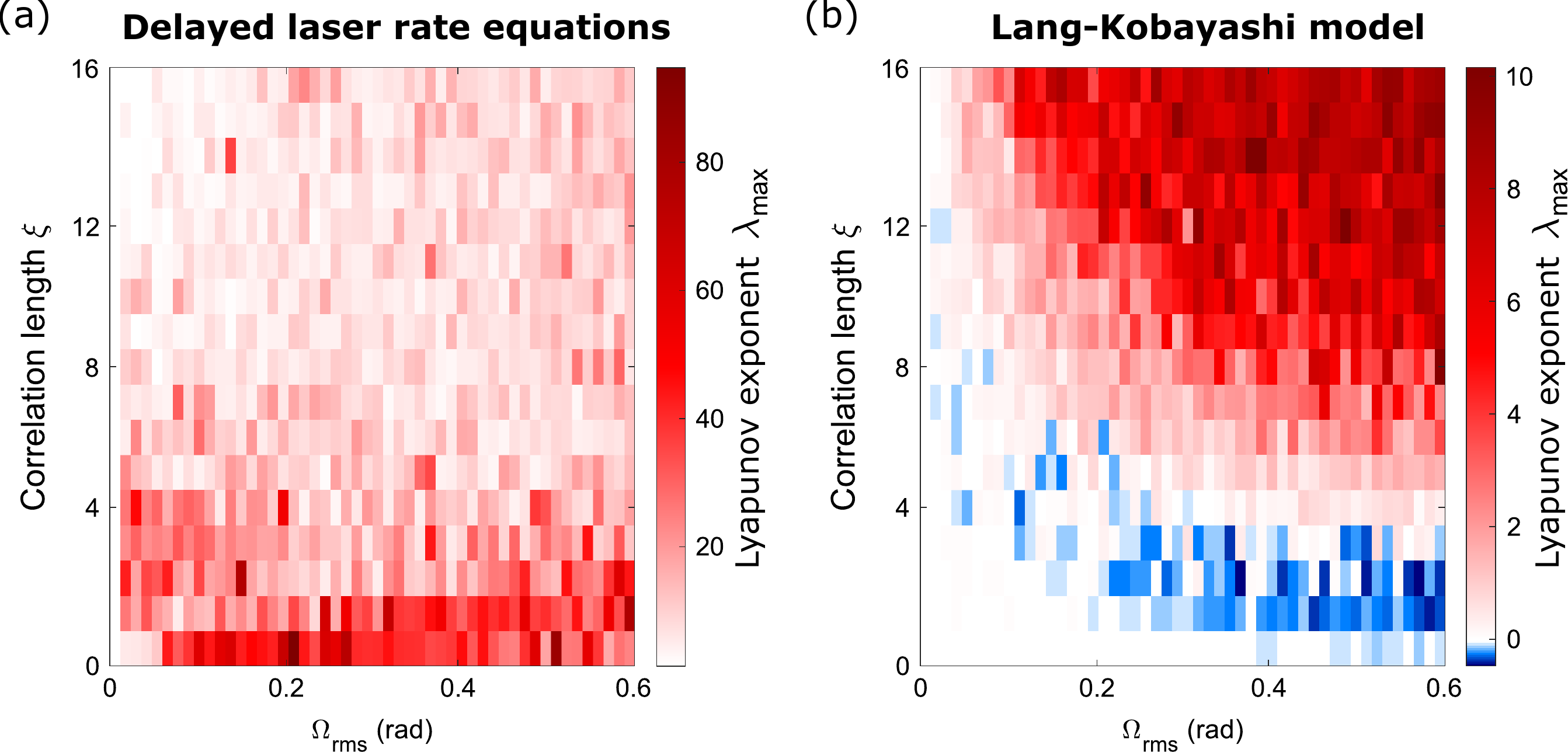}
\rev{\caption{Impact of correlations in parameter disorder in delayed and non-delayed laser models.  Lyapunov exponent $\lambda_{\rm max}$ as a function of the disorder level $\Omega_{\rm{rms}}$ and its correlation length $\xi$.
The left panel shows results for the delayed LRE model, while the right panel shows the corresponding results for the LK model.
The correlation between the detuning at two sites $(i,j)$ and $(i',j')$ is given by a Gaussian decay with distance  $\langle \Omega_{ij} ,\Omega_{i'j'} \rangle \;=\; 
\left(\Omega_{\mathrm{rms}}\right)^{2} 
\exp\!\left[
-\frac{(i-i')^{2} + (j-j')^{2}}{\xi^{2}}
\right],$ where the lasers are coupled as a $5\times 5$ grid ($A_{ij}=1$ if sites $i$ and $j$ are nearest neighbors, and $0$ otherwise). 
%
%
Each cell of the heat map corresponds to an average over 10 disorder realizations, with the same parameter settings as Fig.~\ref{fig:LKvsLREtau0}.
}
\vspace{-0.5 cm}
\label{fig:HeatMaps}}
\end{figure}

\medskip
\textbf{Laser rate equations.} 
Now, we extend our analysis to the LRE model \cite{rogister2004Power, pando2024synchronization}, which is described by the following $3M$ equations:
\begin{equation}
\begin{aligned}
    \dot{r}_j&=\frac{1}{\tau_c}\left(G_j-\gamma\right) r_j+ \frac{\kappa}{\tau_c}\sum_{k=1}^M {A}_{j k} r_k \cos \left(\phi_k-\phi_j\right),
    \\
    \dot{\phi}_j&= \omega_j+ \frac{\kappa}{\tau_c}\sum_{k=1}^M {A}_{j k} \frac{r_k}{r_j} \sin \left(\phi_k-\phi_j\right),
    \\
    \dot{G}_j&= \frac{1}{\tau_f}\left(J_0-G_j \left({sr_j^2}+1\right)\right),
\end{aligned}
   \label{LRE_eqs_polar}
\end{equation}

\noindent
for $j = 1,\ldots,M$,
where $E_j(t) = r_j(t) e^{i\phi_j(t)}$ is the complex electric field and $G_j(t)$ is the medium gain of laser $j$. As in the LK model, $\gamma$ denotes the cavity loss, $\omega_j$ is the frequency detuning of laser $j$, $J_0$ is the pump current, $s$ is the gain saturation coefficient, $\kappa$ is the global coupling strength, and ${A}=(A_{jk})$ is the adjacency matrix. The timescale constants $\tau_c = 13.3 \,{\rm ns}$ and $\tau_f = 13.3\,\mu{\rm s}$ are determined by the cavity round-trip time and the fluorescence time of the upper lasing level, respectively. \rev{Both the LRE and LK models describe class B lasers \cite{paoli1988statistical}, in which the polarization relaxes much faster than the other state variables. Despite this similarity, the LRE are often used as a generic model for coupled class B lasers and can be applied to a broader range of gain media. In contrast, the LK model is specifically tailored for semiconductor lasers, where phase-amplitude coupling is explicitly included through the Kramers-Kronig relations, as represented by the linewidth enhancement factor $\frac{\alpha_j}{2}(G_j-\gamma)$ in Eq. (1). Moreover, in semiconductor laser theory, the carrier density is the fundamental quantity evolving from the rate equation, from which the material gain $G(N,r)$ is computed as a function of the carrier number $N$ and the intensity $r^2$. In the LRE model, $G_j$ itself is taken as the dynamical variable, which is a more phenomenological approach that skips the microscopic $N–G$ relationship.}
%
More importantly, the LRE model does not include time-delay coupling.

Similarly to the results from the non-delayed LK model, introducing disorder to the frequency detuning $\omega_j \sim \mathcal N(0,\sigma_\omega^2)$, $\forall j$, in the LRE model leads to instability for increasing 
$\sigma_{\omega}$ (Fig.~\ref{fig:LKvsLREtau0}(d)).
This result is consistent with the conclusions drawn in Ref.~\cite{pando2024synchronization}.
To evaluate the impact of time-delay coupling, we include time delay in the LRE model by setting the coupling terms in Eq.~\eqref{LRE_eqs_polar} to $A_{jk}r_k(t-\tau)\cos(\phi_k(t-\tau)-\phi_j(t))$ and $A_{jk}\frac{r_k(t-\tau)}{r_j(t)}\sin(\phi_k(t-\tau)-\phi_j(t))$. Surprisingly, despite the similarities between the time-delay LK model and the time-delay LRE model, disorder also \textit{does not} promote synchronization in the time-delay LRE model (Fig.~\ref{fig:LKvsLREtau0}(c)). 
\rev{Inspired by Ref.~\cite{pando2024synchronization}, we further investigated the role of spatial correlations in the frequency-detuning disorder. Figure~\ref{fig:HeatMaps} shows the largest Lyapunov exponent as a function of correlation length and disorder strength in a $5\times 5$ laser grid. 
For the delayed LK model, we observe that a small degree of correlation can promote stability, further supporting the results in Ref.~[41]. Nevertheless, for the delayed LRE model, correlations remain insufficient to promote stability.} 

This analysis highlights that conclusions drawn from one laser model cannot be directly generalized to another.
To gain further insight into why the LRE model does not benefit from disorder-promoted synchronization, we explore an MSF analysis to establish the relationship between the stability measure, time delays, and self-dynamics of the LRE.

\medskip
\textbf{Master stability function analysis of the time-delay LRE model.}
First, we note that there is a clear timescale separation in the LRE between the dynamics of the electric field and the gain since $\tau_f \gg \tau_c$. This allows us to reduce the system to a set of $2M$ equations by using a quasi-steady-state approximation in which $\dot G_j \approx 0$, yielding $G_j = \frac{J_0}{1+sr_j^2}$.
Considering the identical synchronous solution $E_j(t) = r^* e^{i\Omega t}$, under the assumption that all lasers are identical (i.e., $\omega_j = \omega$, $\forall j$), we obtain the following $2M$ transcendental equations:
\begin{equation}
\begin{aligned}
    0&=\frac{1}{\tau_c}\left(\frac{J_0}{1+sr_j^2(t)}-\gamma\right) r_j(t)+ \frac{\kappa}{\tau_c}\sum_{k=1}^M {A}_{j k} r_k(t-\tau) \cos \left(\phi_k(t-\tau)-\phi_j(t)\right),
    \\
    \Omega&= \omega + \frac{\kappa}{\tau_c}\sum_{k=1}^M {A}_{j k} \frac{r_k(t-\tau)}{r_j(t)} \sin \left(\phi_k(t-\tau)-\phi_j(t)\right).
\end{aligned}
   \label{LRE_trans}
\end{equation}

\noindent
By solving these equations, we determine the stationary values $r_j^*$ and $\Omega$, for $j=1,\ldots,M$. Assuming $\omega = 0$ (without loss of generality), it follows that $\Omega = 0$ is a solution of Eq.~\eqref{LRE_trans}. In contrast, $\Omega$ is always non-zero in the LK model due to the phase-amplitude coupling. As we show next, this difference has direct implications for the linearized dynamics of the LRE model.

We now apply the MSF theory from Ref.~\cite{choe2010controlling} developed for time-delay systems. Importantly, this theory holds for an adjacency matrix $A$ satisfying the assumption that all nodes have a common indegree $d$. The MSF analysis allows us to characterize the stability of the time-delay LRE model using the following variational equation:
\begin{equation}
    \dot \zeta_j(t) = \underbrace{\left(\operatorname{D}^{(0)} \textbf{f}+\kappa d^* \operatorname{D}^{(0)}\textbf{h} \right)}_{{J}_1} \bm\zeta_j(t)+\underbrace{\left(\kappa\nu \operatorname{D}^{(\tau)}\textbf{h}\right)}_{{J}_2} \,\, \bm\zeta_j(t-\tau).
\end{equation}

\noindent
Here, $\nu$ is an eigenvalue of $A$, and $J_1$ and $J_2$ are $2\times 2$ time-independent matrices determined by the self-dynamics $\textbf{f}$ and coupling term \textbf{h} of the time-delay LRE model. It follows that
\begin{equation}
    \operatorname{D}^{(\tau)}\textbf{h} =
\begin{bmatrix}
\cos(-\Omega\tau) & -r^*\sin(-\Omega\tau)\\
\frac{1}{r^*}\sin(-\Omega\tau) & \cos(-\Omega\tau)
\end{bmatrix} = I_2.
\end{equation}

\noindent 
Substituting $J_2 = \kappa\nu I_2$ in the characteristic equation (6), we have that
\begin{equation}
\det\left(J_1 - (\lambda_{\ell} - \kappa\nu e^{-\lambda_{\ell}\tau} )I_2 \right) = 0.
\end{equation}

\noindent
The stability exponents $\lambda_{\ell}$ of the time-delay LRE model is thus directly determined by the eigenvalues $\tilde\lambda_{\ell}$ of matrix $J_1$ according to the relation $\lambda_{\ell} - \kappa\nu e^{-\lambda_{\ell}\tau} = \tilde\lambda_{\ell}$. For small $\tau$, we consider the Taylor series expansion $e^{-\lambda_{\ell}\tau} \approx 1 - \lambda_{\ell}\tau$, yielding the relation
\begin{equation}
    \lambda_{\ell} = \frac{\tilde{\lambda}_{\ell} + \kappa\nu}{1+\kappa\nu\tau}.
\end{equation}

\noindent
Therefore, increasing the coupling strength $\kappa$ shifts the eigenvalues of $J_1$ rightward in the complex plane, leading to instability of the synchronous state. In contrast, increasing the time delay $\tau$ in the LRE model can only reduce the magnitude of the eigenvalues of $J_1$, and is thus insufficient to cause the synchronous state to shift from stable to unstable. This behavior is significantly different from the one exhibited by the LK model, where increasing either the coupling strength $\kappa$ or the time delay $\tau$ leads to instability (cf.\ Fig.~\ref{fig:ArnoldTonges}).


\rev{\section{Comparison between laser classes}}
\label{sec.ClassesComparison}

\rev{Given the broad diversity of laser systems, ranging from their construction and choice of gain medium to the characteristic time scales, it is essential to distinguish their dynamical classes and typical applications. In this section, we outline the main differences among laser classes and discuss how adding or removing degrees of freedom in both the LRE and the LK model influences the role of heterogeneity in stability.}

\begin{table*}
    \centering
\begin{footnotesize}
    \begin{tabular}{ccc}
    \toprule[1.5pt]
        \textbf{Laser class} & \textbf{Laser rate equations} & \textbf{Lang-Kobayashi model}\\
        & \rev{(weak phase-amplitude coupling)} & \rev{(strong phase-amplitude coupling)} \\
    \toprule[1.5pt]
    \addlinespace[3pt]  
\makecell{\textbf{Class A} \\ (reduction)} & 
$\begin{aligned} 
& \dot{E}_j=\frac{1}{\tau_c}(G_j - \gamma) E_j+i \omega_j E_j +\frac{\kappa_j}{\tau_c} \sum_k A_{j k} E_k(t-\tau)  
\end{aligned}$                  
& 
$\begin{aligned}
& \dot{E}_j=\frac{1+i\alpha_j}{2}(G_j(t) -\gamma)E_j +i \omega_j E_j+k_j \sum_k A_{j k} E_k(t-\tau)
\end{aligned}$\\
\addlinespace[3pt]  
\midrule[0.5pt]
\addlinespace[3pt]  
\textbf{Class B}   & 
$\begin{aligned}
& \dot{E}_j=\rev{\frac{1}{\tau_c}}(G_j -\gamma)E_j +i \omega_j E_j+\frac{\kappa_j}{\tau_c} \sum_k A_{j k} E_k(t-\tau)\\ 
& \dot{G}_j= \frac{1}{\tau_f}\left(J_0-G_j \left({s\left|E_j\right|^2}+1\right)\right)
\end{aligned}$
&
$\begin{aligned}
& \dot{E}_j=\frac{1+i\alpha_j}{2}(G_j(t) -\gamma)E_j +i \omega_j E_j+\kappa_j \sum_k A_{j k} E_k(t-\tau) \\
& \dot{N}_j=J_0-\gamma_n N_j-G_j(t)\left|E_j\right|^2
\end{aligned}$ \\
\addlinespace[3pt]  
\midrule[0.5pt]
\addlinespace[3pt]  
\makecell{\textbf{Class C} \\ \rev{(extension)}} & 
$\begin{aligned} 
 & \dot{E}_j=-\frac{\gamma}{\tau_c} E_j+\frac{1}{\tau_c} P_j+i \omega_jE_j+\frac{\kappa}{\tau_c} \sum_k A_{j k} E_k(t-\tau) \\ & \dot{P}_j=\gamma_\perp\left(-P_j+G_j E_j\right) \\ & 
 \dot{G}_j= \frac{1}{\tau_f}\left(J_0-G_j \left({s\left|E_j\right|^2}+1\right)\right)
\end{aligned}$ & 
$\begin{aligned}
& \dot{E}_j= -\frac{\gamma}{2} E_j - \rev{P_j} + i \omega_{j} E_j +\kappa_j \sum_k A_{j k} E_k(t-\tau) \\
& \dot{P}_j= -(\gamma_\perp + \rev{i \Delta})P_j+ G_j(t) E_j\\
& \dot{N}_j=J_0-\gamma_n N_j-G_j(t)\left|E_j\right|^2
\end{aligned}$\\
\addlinespace[3pt]  
\toprule[1.5pt]
\addlinespace[3pt]  
\end{tabular} 
\end{footnotesize}
    \vspace{-0.3cm}
    \rev{\caption{\label{tab.ClassesModels} Dynamical equations for different laser classes of LRE and LK models.} }
\vspace{-0.4cm}
\end{table*}

\medskip
\rev{\textbf{Class A lasers.} 
In class A laser models, the dynamics represent systems where both $P$ and $N$ decay much faster than photons. Therefore, only the electric field $E$ remains as a dynamical variable, and the polarization and the carrier number can be adiabatically eliminated \cite{haken1985laser}. These models can describe dye lasers, some gas lasers in high-pressure regimes \cite{sargent1974laser}, and solid-state lasers with very fast carrier dynamics compared to cavity lifetime \cite{erneux2010laser}. We note that semiconductors are rarely described by class A models, since carriers are slow compared to photons.}

\medskip
\rev{\textbf{Class B lasers.} 
In class B laser models, the polarization is fast and can be eliminated, but photon and carrier lifetimes are comparable, leaving the electric field $E$ and the carrier number $N$ as the dynamical variables. This is the most common class of lasers in practice. Among the class B models are the laser rate equations and the LK model, where the first is a more general phenomenological description of class B lasers, whereas the LK model is specifically tailored for semiconductor lasers and incorporates features such as phase-amplitude coupling and time-delay feedback. 
Class B lasers also exhibit relaxation oscillations, self-pulsing, and chaos under modulation or feedback.
Examples of lasers that can be described by this model include semiconductor diode lasers (GaAs, InP, etc.), which are the focus of our study, and most solid-state lasers (Nd:YAG, fiber lasers).}

\medskip
\rev{\textbf{Class C lasers.} 
In class C laser models, all three variables $E$, $P$, and $N$ have comparable time scales and thus must be considered as state variables. Therefore, the equations are given by the full Maxwell–Bloch system, with no adiabatic elimination. This model allows for richer dynamics, which can support Rabi oscillations \cite{lingnau2019class}, polarization beats, and chaos \cite{lugiato1983breathing}. This class often describes gas lasers with long-lived atomic coherences (e.g., argon-ion, $\mathrm{CO_2}$, He–Ne lasers in low-pressure conditions), some micro or nanolasers \cite{roos2021stabilizing}, and quantum-dot lasers operated near resonance with long dephasing times \cite{lingnau2012failure}. These models are often reduced to Class B for semiconductors because of how short the polarization lifetime is and due to the higher mathematical complexity. 
}

\medskip
The dynamical equations for each laser class are summarized in Table~\ref{tab.ClassesModels}. Class A equations were derived by reducing the LK and LRE class B models through a quasi-steady-state approximation $\dot N_j\approx 0$ that eliminates the carrier number dynamics (or the gain dynamics  $\dot G_j\approx 0$ in the case of the LRE). \rev{Class C equations were obtained by extending both models to incorporate the polarization dynamics according to the Maxwell-Bloch equations \eqref{eq:Max-Bloch} \cite{arecchi1984deterministic}. 
As discussed in Section~\ref{sec.LEfortau0}, the gain $G_j(t)$ is directly modeled as a state variable in the LRE model, whereas in the LK model it represents a function of the carrier number and the field amplitude (i.e., $G_j(t)=g \frac{N_j(t)-N_{0}}{1+s|E_j|^2(t)}$).}

\rev{In the LK class C model, the $\alpha$ factor is no longer a valid approximation for the phase–amplitude coupling, as the ratio of refractive index change to gain becomes time- and frequency-dependent when polarization evolves according to its own timescale \cite{lingnau2012failure}. In the LK class B model, the linewidth enhancement factor is given by $\alpha = \frac{\Delta}{\gamma_\perp}$, derived through the adiabatic elimination of $P$ \cite{henry2003theory}. Consequently, the frequency detuning in the class C model relates to that of the class B model through $\omega^{(\rm class \,\, C)} = \omega^{(\rm class \,\, B)} - \frac{1}{2}\alpha\gamma$. Even though classes A and C of the LK model are not typically used in practice for semiconductor diode lasers (since they are generally well described by class B dynamics), we include them in this analysis for completeness.}


\begin{figure}[t]
\centering
\includegraphics[width=0.9\linewidth]{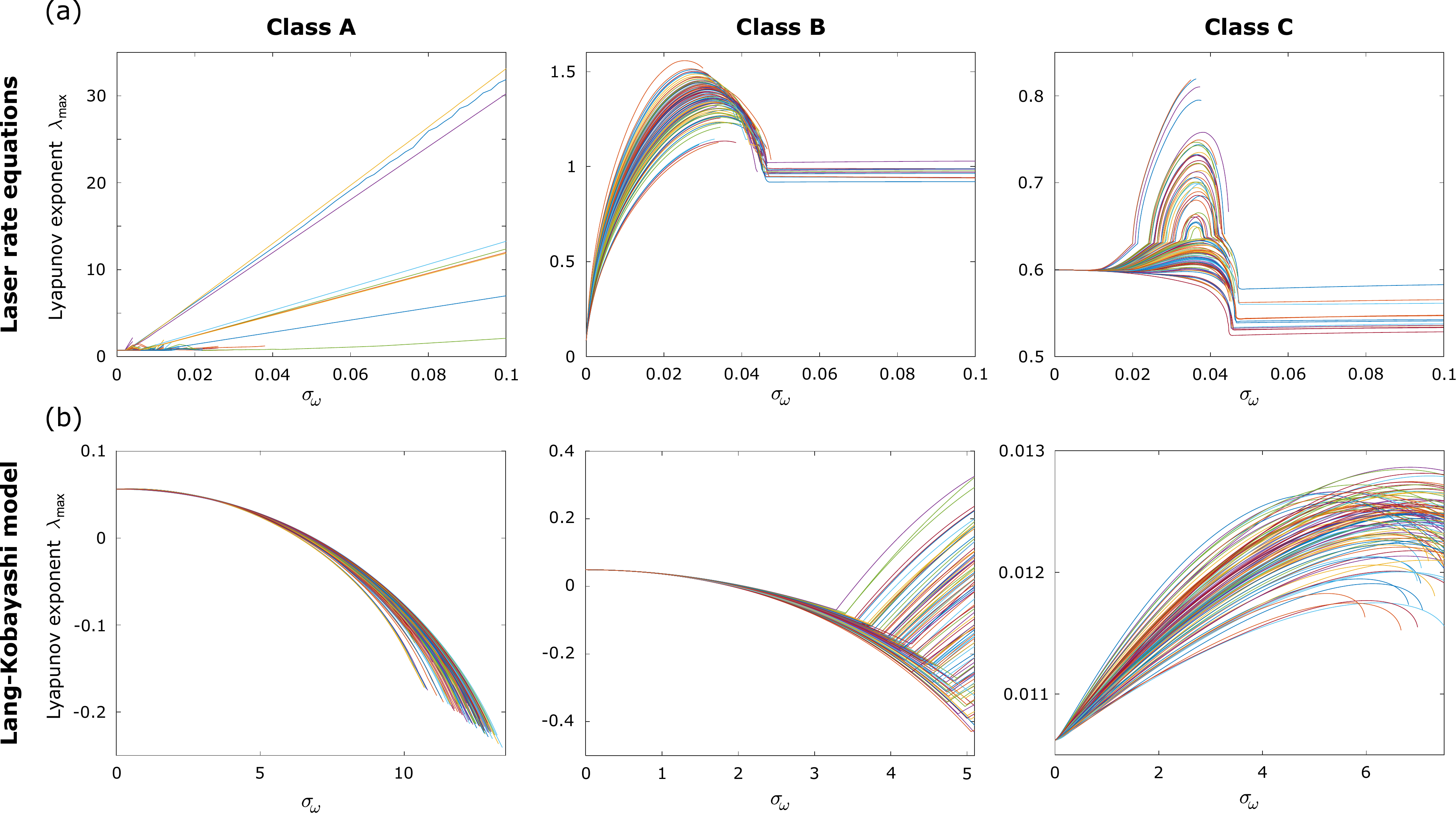}
\rev{\caption{Impact of parameter disorder in different classes of LRE and LK models.  Lyapunov exponent $\lambda_{\rm max}$ as a function of $\sigma_\omega$ for each of the 100 realizations in a decaying network with 10 lasers. The panels present the stability curves for the LRE (left) and LK (right) models across different laser classes: class A (left), class B (middle), and class C (right). The same parameter settings are the same as in Fig.~\ref{fig:LKvsLREtau0}; for the class C models, the polarization decay rate is set to $\gamma_\perp = 0.5$ for the LRE, \rev{and $\gamma_\perp = 0.1$, $\Delta = 0.5$ for the LK model}.
}
\label{fig:LRE_LK_Classes}}
\end{figure}

\begin{figure}[t]
\centering
\includegraphics[width=0.65\linewidth]{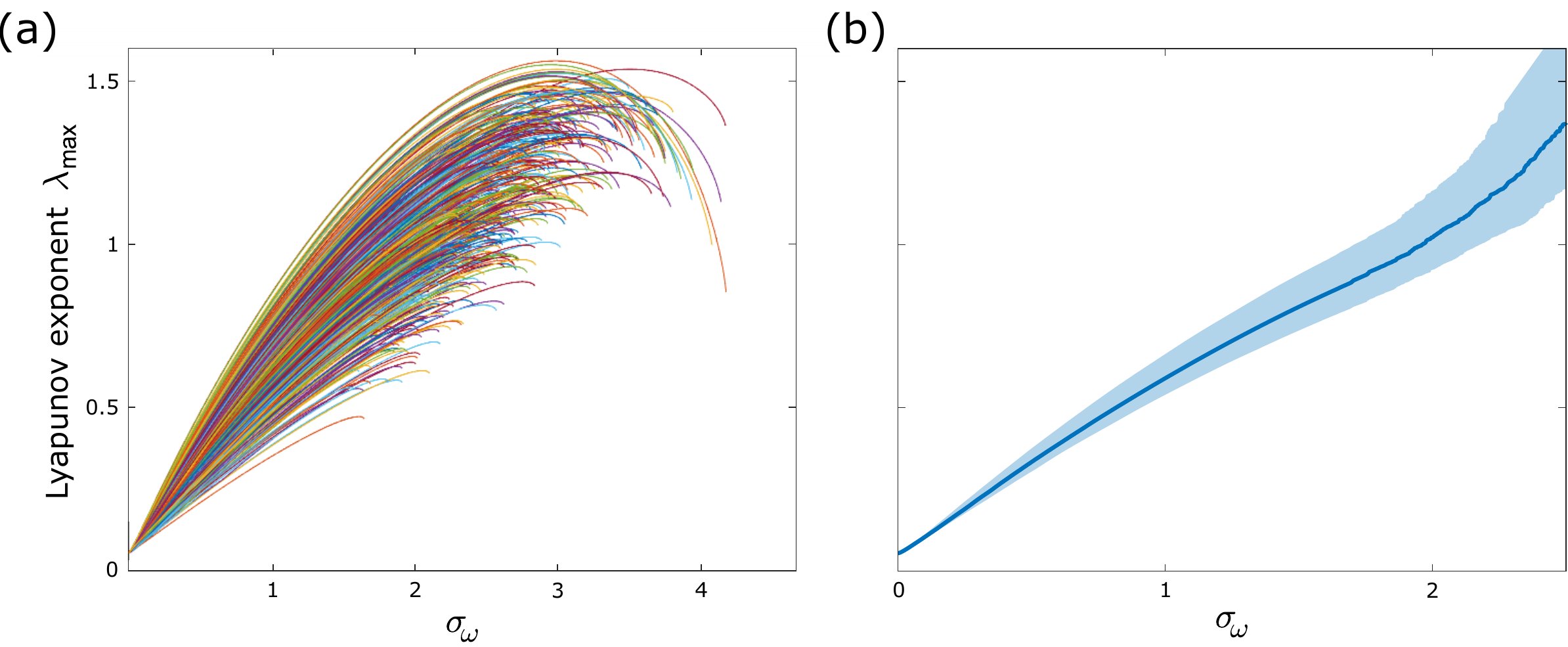}
\rev{\caption{Impact of parameter disorder in the absence of saturation.  (a) Lyapunov exponent $\lambda_{\rm max}$ as a function of $\sigma_\omega$ for each of the 100 realizations in a decaying network with 10 lasers with $s=0$ and $\kappa =0.2\, \rm{ns}^{-1}$. (b) Median (solid line) with first and third quartiles (shaded areas) of the Lyapunov exponent $\lambda_{\rm max}$ across the independent disorder realizations.
}
\vspace{-0.4 cm}
\label{fig:MtleNoSat}}
\end{figure}

\medskip
\rev{\textbf{Comparison between different degrees of freedom in LRE and LK models.}} 
\rev{To examine the role of the dynamical degrees of freedom on disorder-promoted synchronization, Fig. \ref{fig:LRE_LK_Classes} shows the effect of introducing disorder into the frequency detuning $\omega_j$ in each of the laser classes, including both the LRE and LK formulations.
For the LRE, Fig.~\ref{fig:LRE_LK_Classes}(a) shows that the system remains unstable across all cases, although the stability behavior depends strongly on the number of dynamical variables. In class A and B lasers, disorder is strictly detrimental, yielding larger $\lambda_{\rm max}$ relative to the homogeneous case. Extending the model to class C reveals a different trend: although the system is intrinsically more unstable, several realizations display improved stability with increasing heterogeneity. Nevertheless, stability is never fully achieved.
For the LK model, Fig.~\ref{fig:LRE_LK_Classes}(b) shows that the reduction to class A also exhibits disorder-promoted stability. In these cases, higher levels of disorder are generally needed, and bifurcations always occur within stable regimes. \rev{In contrast, for the class C extension of the LK model, heterogeneity no
longer improves stability.}}


\rev{The comparison between the LRE and LK models highlights the crucial role of model structure, particularly phase-amplitude coupling and the phenomenological treatment of gain, in determining whether disorder can enhance or suppress stability.
Since a key distinction between the LK and LRE laser types lies in the nature and dynamics of their gain media, we investigated the impact of heterogeneity in a system without gain saturation, by setting $s = 0$. In this case, Fig.~\ref{fig:MtleNoSat} shows that the beneficial impact of disorder vanishes, highlighting the essential role of nonlinearity in enabling disorder-promoted synchronization. 
}

\end{document}